\documentstyle[11pt,subfigure,graphicx,color,cancel,cite,ulem,framed,hyperref,url,multirow,lmodern,fontenc,amsmath]{article}
 \hypersetup{linktocpage}
 \textwidth16.5cm\begin{tiny}

\end{tiny}
\newcommand{\newc}{\newcommand}

\def\Ord{\lower .7ex\hbox{$\;\stackrel{\textstyle <}{\sim}\;$}}
\def\OOrd{\lower .7ex\hbox{$\;\stackrel{\textstyle >}{\sim}\;$}}

\newc{\order}{{\cal O}}

\def\lum             {{\cal L}}

\def\M               {{\cal M}}

\newc{\be}{\begin{equation}}
\newc{\ee}{\end{equation}}
\newc{\br}{\begin{eqnarray}}
\newc{\er}{\end{eqnarray}}
\newc{\ba}{\begin{array}}
\newc{\ea}{\end{array}}
\newc{\bi}{\begin{itemize}}
\newc{\ei}{\end{itemize}}
\newc{\bn}{\begin{enumerate}}
\newc{\en}{\end{enumerate}}
\newc{\bc}{\begin{center}}
\newc{\ec}{\end{center}}
\newc{\ul}{\underline}
\newc{\ra}{\rightarrow}
\newc{\lra}{\longrightarrow}
\newc{\wt}{\widetilde}
\newc{\til}{\tilde}

\newc{\wh}{\widehat}
\newc{\ti}{\times}
\newc{\Dir}{\kern -6.4pt\Big{/}}
\newc{\Dirin}{\kern -10.4pt\Big{/}\kern 4.4pt}
\newc{\DDir}{\kern -10.6pt\Big{/}}
\newc{\DGir}{\kern -6.0pt\Big{/}}
\newc{\sig}{\sigma}
\newc{\sigmalstop}{\sig_{\lstoppair}}
\newc{\Sig}{\Sigma}  
\newc{\del}{\delta}
\newc{\Del}{\Delta}
\newc{\lam}{\lambda}
\newc{\Lam}{\Lambda}
\newc{\gam}{\gamma}
\newc{\Gam}{\Gamma}
\newc{\eps}{\epsilon}
\newc{\Eps}{\Epsilon}
\newc{\kap}{\kappa}
\newc{\Kap}{\Kappa}
\newc{\modulus}[1]{\left| #1 \right|}
\newc{\eq}[1]{(\ref{eq:#1})}
\newc{\eqs}[2]{(\ref{eq:#1},\ref{eq:#2})}
\newc{\etal}{{\it et al.}\ }
\newc{\ibid}{{\it ibid}.}
\newc{\ibidem}{{\it ibidem}.}
\newc{\eg}{{\it e.g.}\ }
\newc{\ie}{{\it i.e.}\ }

\newc{\nonum}{\nonumber}
\newc{\lab}[1]{\label{eq:#1}}
\newc{\dpr}[2]{({#1}\cdot{#2})}
\newc{\lt}{\stackrel{<}}
\newc{\gt}{\stackrel{>}}
\newc{\lsimeq}{\stackrel{<}{\sim}}
\newc{\gsimeq}{\stackrel{>}{\sim}}
\def\lsim{\buildrel{\scriptscriptstyle <}\over{\scriptscriptstyle\sim}}
\def\gsim{\buildrel{\scriptscriptstyle >}\over{\scriptscriptstyle\sim}}
\def\lapp{\mathrel{\rlap{\raise.5ex\hbox{$<$}}
                    {\lower.5ex\hbox{$\sim$}}}}
\def\gapp{\mathrel{\rlap{\raise.5ex\hbox{$>$}}
                    {\lower.5ex\hbox{$\sim$}}}}
\newc{\half}{\frac{1}{2}}

\newc{\bQ}{\ol{Q}}
\newc{\dota}{\dot{\alpha }}
\newc{\dotb}{\dot{\beta }}
\newc{\dotd}{\dot{\delta }}
\newc{\nindnt}{\noindent}

\newc{\matth}{\mathsurround=0pt}
\def\ML{\ifmmode{{\mathaccent"7E M}_L}
             \else{${\mathaccent"7E M}_L$}\fi}
\def\MR{\ifmmode{{\mathaccent"7E M}_R}
             \else{${\mathaccent"7E M}_R$}\fi}

\newc{\mr}{\mathrm}

\newc{\siminf}{\mbox{$_{\sim}$ {\small {\hspace{-1.em}{$<$}}}    }}
\newc{\simsup}{\mbox{$_{\sim}$ {\small {\hspace{-1.em}{$>$}}}    }}

\newc {\Zboson}{{\mathrm Z}^{0}}
\newc{\thetaw}{\theta_W}
\newc{\mbot}{{m_b}}
\newc{\mtop}{{m_t}}
\newc{\sm}{${\cal {SM}}$}
\newc{\as}{\alpha_s}
\newc{\aem}{\alpha_{em}}

\newc{\ppbar}{\mbox{$p\ol{p}$}}
\newc{\bbbar}{\mbox{$b\ol{b}$}}
\newc{\ccbar}{\mbox{$c\ol{c}$}}
\newc{\ttbar}{\mbox{$t\ol{t}$}}
\newc{\eebar}{\mbox{$e\ol{e}$}}
\newc{\zzero}{\mbox{$Z^0$}}

\newc{\wplus}{\mbox{$W^+$}}
\newc{\wminus}{\mbox{$W^-$}}
\newc{\ellp}{\ell^+}
\newc{\ellm}{\ell^-}
\newc{\elp}{\mbox{$e^+$}}
\newc{\elm}{\mbox{$e^-$}}
\newc{\elpm}{\mbox{$e^{\pm}$}}
\newc{\qbar}     {\mbox{$\ol{q}$}}

\newc{\Ebar}{{\bar E}}
\newc{\Dbar}{{\bar D}}
\newc{\Ubar}{{\bar U}}
\newc{\susy}{{{SUSY}}}
\newc{\msusy}{{{M_{SUSY}}}}

\def\photino{\ifmmode{\mathaccent"7E \gam}\else{$\mathaccent"7E \gam$}\fi}
\def\taugluino{\ifmmode{\tau_{\mathaccent"7E g}}
             \else{$\tau_{\mathaccent"7E g}$}\fi}
\def\mphotino{\ifmmode{m_{\mathaccent"7E \gam}}
             \else{$m_{\mathaccent"7E \gam}$}\fi}
\newc{\gl}   {\mbox{$\wt{g}$}}
\newc{\mgl}  {\mbox{$m_{\gl}$}}

\def \chonep {{\wt\chi_1^+}}

\def \ch2p {{\wt\chi_2^+}}
\def \chonem {{\wt\chi_1^-}}
\def \ch2m {{\wt\chi_2^-}}

\def \chonepm{{\wt\chi_1}^{\pm}}

\def \mchonepm{m_{\chonepm}}

\def \chtwopm{{\wt\chi_2}^{\pm}}

\newc{\dmchi}{\Delta m_{\wt\chi}}

\def \lspone{\wt\chi_1^0}
\def \mlspone{m_{\lspone}}
\def \lsptwo{\wt\chi_2^0}
\def \mlsptwo{m_{\lsptwo}}
\def \lspthree{\wt\chi_3^0}
\def \mlspthree{m_{\lspthree}}
\def \lspfour{\wt\chi_4^0}

\newc{\sele}{\wt{\mathrm e}}
\newc{\sell}{\wt{\ell}}

\def \stauone{\wt\tau_1}
\def \stauonepm{{\wt\tau_1}^\pm}
\def \mstauone{m_{\stauone}}
\def \snu{\wt{\nu}}
\def \snutau{\wt{\nu}_{\tau}}
\def \nutau{{\nu}_{\tau}}                  

\newc{\snue}     {\mbox{$ \wt{\nu_e}$}}
\newc{\smu}{\wt{\mu}}
\newc{\stau}{\wt{\tau}}
\newc {\nuL} {\wt{\nu}_L}
\newc {\nuR} {\wt{\nu}_R}
\newc {\snub} {\bar{\wt{\nu}}}
\newc {\eL} {\wt{e}_L}
\newc {\eR} {\wt{e}_R}
\def \slepl{\wt{l}_L}

\def \slepr{\wt{l}_R}

\def \stau{\wt\tau}

\def \slepton{\wt\ell}

\def \sq{\wt{q}}

\newc{\msqot}  {\mbox{$m_(\sq_{1,2} )$}}
\newc{\sqbar}    {\mbox{$\bar{\wt{q}}$}}
\newc{\ssb}      {\mbox{$\squark\ol{\squark}$}}
\newc {\qL} {\wt{q}_L}
\newc {\qR} {\wt{q}_R}
\newc {\uL} {\wt{u}_L}
\newc {\uR} {\wt{u}_R}
\def \ul{\wt{u}_L}

\newc {\dL} {\wt{d}_L}
\newc {\dR} {\wt{d}_R}
\newc {\cL} {\wt{c}_L}
\newc {\cR} {\wt{c}_R}
\newc {\sL} {\wt{s}_L}
\newc {\sR} {\wt{s}_R}
\newc {\tL} {\wt{t}_L}
\newc {\tR} {\wt{t}_R}
\newc {\stb} {\ol{\wt{t}}_1}
\newc {\sbot} {\wt{b}_1}
\newc {\msbot} {m_{\sbot}}
\newc {\sbotb} {\ol{\wt{b}}_1}
\newc {\bL} {\wt{b}_L}
\newc {\bR} {\wt{b}_R}

\newc{\csquark}  {\mbox{$\wt{c}$}}
\newc{\csquarkl} {\mbox{$\wt{c}_L$}}
\newc{\mcsl}     {\mbox{$m(\csquarkl)$}}

\newc {\stopl}         {\wt{\mathrm{t}}_{\mathrm L}}
\newc {\stopr}         {\wt{\mathrm{t}}_{\mathrm R}}
\newc {\stoppair}      {\wt{\mathrm{t}}_{1}
\bar{\wt{\mathrm{t}}}_{1}}
\def \lstop{\wt{t}_{1}}

\def \hstop{\wt{t}_{2}}

\def \mlstop{m_{\lstop}}
\def \mhstop{m_{\hstop}}
\def \lstoppair{\lstop\lstop^*}

\newc{\tsquark}  {\mbox{$\wt{t}$}}
\newc{\ttb}      {\mbox{$\tsquark\ol{\tsquark}$}}
\newc{\ttbone}   {\mbox{$\tsquark_1\ol{\tsquark}_1$}}

\newc{\mix}{\theta_{\wt t}}
\newc{\cost}{\cos{\theta_{\wt t}}}
\newc{\sint}{\sin{\theta_{\wt t}}}
\newc{\costloop}{\cos{\theta_{\wt t_{loop}}}}

\newc{\mixsbot}{\theta_{\wt b}}

\newc{\tb}{\tan\beta}
\newc{\cb}{\cot\beta}
\newc{\vev}[1]{{\left\langle #1\right\rangle}}

\newc{\mhalf}{m_{1/2}}
\newc{\mzero} {\mbox{$m_0$}}
\newc{\azero} {\mbox{$A_0$}}

\newc{\lb}{\lam}
\newc{\lbp}{\lam^{\prime}}
\newc{\lbpp}{\lam^{\prime\prime}}
\newc{\rpv}{{\not \!\! R_p}}
\newc{\rpvm}{{\not  R_p}}
\newc{\rp}{R_{p}}
\newc{\rpmssm}{{RPC MSSM}}
\newc{\rpvmssm}{{RPV MSSM}}

\newc{\sbyb}{S/$\sqrt B$}
\newc{\pelp}{\mbox{$e^+$}}
\newc{\pelm}{\mbox{$e^-$}}
\newc{\pelpm}{\mbox{$e^{\pm}$}}
\newc{\epem}{\mbox{$e^+e^-$}}
\newc{\lplm}{\mbox{$\ell^+\ell^-$}}

\def\Ecm{\ifmmode{E_{\mathrm{cm}}}\else{$E_{\mathrm{cm}}$}\fi}
\newc{\rts}{\sqrt{s}}
\newc{\rtshat}{\sqrt{\hat s}}
\newc{\gev}{\,GeV}
\newc{\mev}{~{\rm MeV}}
\newc{\tev}  {\mbox{$\;{\rm TeV}$}}
\newc{\gevc} {\mbox{$\;{\rm GeV}/c$}}
\newc{\gevcc}{\mbox{$\;{\rm GeV}/c^2$}}
\newc{\intlum}{\mbox{${ \int {\cal L} \; dt}$}}
\newc{\call}{{\cal L}}
\def \met  {\mbox{${E\!\!\!\!/_T}$}}

\newc{\ptmiss}{/ \hskip-7pt p_T}

\newc{\PT}{\mbox{$p_T$}}
\newc{\ET}{\mbox{$E_T$}}
\newc{\dedx}{\mbox{${\rm d}E/{\rm d}x$}}
\newc{\ifb}{\mbox{${\rm fb}^{-1}$}}
\newc{\ipb}{\mbox{${\rm pb}^{-1}$}}
\newc{\pb}{~{\rm pb}}
\newc{\fb}{~{\rm fb}}
\newc{\ycut}{y_{\mathrm{cut}}}
\newc{\chis}{\mbox{$\chi^{2}$}}

\def \jet(s){\emph{jet(s) }}

\newc{\mpl}{M_{\rm Pl}}
\newc{\mgut}{M_{GUT}}
\newc{\mw}{M_{W}}
\newc{\mweak}{M_{weak}}
\newc{\mz}{M_{Z}}

\newc{\OPALColl}   {OPAL Collaboration}
\newc{\ALEPHColl}  {ALEPH Collaboration}
\newc{\DELPHIColl} {DELPHI Collaboration}
\newc{\XLColl}     {L3 Collaboration}
\newc{\JADEColl}   {JADE Collaboration}
\newc{\CDFColl}    {CDF Collaboration}
\newc{\DXColl}     {D0 Collaboration}
\newc{\HXColl}     {H1 Collaboration}
\newc{\ZEUSColl}   {ZEUS Collaboration}
\newc{\LEPColl}    {LEP Collaboration}
\newc{\ATLASColl}  {ATLAS Collaboration}
\newc{\CMSColl}    {CMS Collaboration}
\newc{\UAColl}    {UA Collaboration}
\newc{\KAMLANDColl}{KamLAND Collaboration}
\newc{\IMBColl}    {IMB Collaboration}
\newc{\KAMIOColl}  {Kamiokande Collaboration}
\newc{\SKAMIOColl} {Super-Kamiokande Collaboration}
\newc{\SUDANTColl} {Soudan-2 Collaboration}
\newc{\MACROColl}  {MACRO Collaboration}
\newc{\GALLEXColl} {GALLEX Collaboration}
\newc{\GNOColl}    {GNO Collaboration}
\newc{\SAGEColl}  {SAGE Collaboration}
\newc{\SNOColl}  {SNO Collaboration}
\newc{\CHOOZColl}  {CHOOZ Collaboration}
\newc{\PDGColl}  {Particle Data Group Collaboration}

\def\issue(#1,#2,#3){{\bf #1}, #2 (#3)}
\def\iss(#1,#2,#3){{\bf #1} (#3) #2}
\def\ASTR(#1,#2,#3){Astropart.\ Phys. \issue(#1,#2,#3)}
\def\AJ(#1,#2,#3){Astrophysical.\ Jour. \issue(#1,#2,#3)}
\def\AJS(#1,#2,#3){Astrophys.\ J.\ Suppl. \issue(#1,#2,#3)}
\def\APP(#1,#2,#3){Acta.\ Phys.\ Pol. \issue(#1,#2,#3)}
\def\JCAP(#1,#2,#3){Journal\ XX. \issue(#1,#2,#3)} 
\def\SC(#1,#2,#3){Science \issue(#1,#2,#3)}
\def\PRD(#1,#2,#3){Phys.\ Rev.\ D \issue(#1,#2,#3)}
\def\PR(#1,#2,#3){Phys.\ Rev.\ \issue(#1,#2,#3)} 
\def\PRC(#1,#2,#3){Phys.\ Rev.\ C \issue(#1,#2,#3)}
\def\NPB(#1,#2,#3){Nucl.\ Phys.\ B \issue(#1,#2,#3)}
\def\NPPS(#1,#2,#3){Nucl.\ Phys.\ Proc. \ Suppl \issue(#1,#2,#3)}
\def\NJP(#1,#2,#3){New.\ J.\ Phys. \issue(#1,#2,#3)}
\def\JP(#1,#2,#3){J.\ Phys.\issue(#1,#2,#3)}
\def\PL(#1,#2,#3){Phys.\ Lett. \issue(#1,#2,#3)}
\def\ZP(#1,#2,#3){Z.\ Phys. \issue(#1,#2,#3)}
\def\ZPC(#1,#2,#3){Z.\ Phys.\ C  \issue(#1,#2,#3)}
\def\PREP(#1,#2,#3){Phys.\ Rep. \issue(#1,#2,#3)}
\def\PRL(#1,#2,#3){Phys.\ Rev.\ Lett. \issue(#1,#2,#3)}
\def\MPL(#1,#2,#3){Mod.\ Phys.\ Lett. \issue(#1,#2,#3)}
\def\RMP(#1,#2,#3){Rev.\ Mod.\ Phys. \issue(#1,#2,#3)}
\def\SJNP(#1,#2,#3){Sov.\ J.\ Nucl.\ Phys. \issue(#1,#2,#3)}
\def\CPC(#1,#2,#3){Comp.\ Phys.\ Comm. \issue(#1,#2,#3)}
\def\IJMPA(#1,#2,#3){Int.\ J.\ Mod. \ Phys.\ A \issue(#1,#2,#3)}
\def\MPLA(#1,#2,#3){Mod.\ Phys.\ Lett.\ A \issue(#1,#2,#3)}
\def\PTP(#1,#2,#3){Prog.\ Theor.\ Phys. \issue(#1,#2,#3)}
\def\RMP(#1,#2,#3){Rev.\ Mod.\ Phys. \issue(#1,#2,#3)}
\def\NIMA(#1,#2,#3){Nucl.\ Instrum.\ Methods \ A \issue(#1,#2,#3)}
\def\EPJC(#1,#2,#3){Eur.\ Phys.\ J.\ C \issue(#1,#2,#3)}
\def\RPP (#1,#2,#3){Rept.\ Prog.\ Phys. \issue(#1,#2,#3)}
\def\PPNP(#1,#2,#3){ Prog.\ Part.\ Nucl.\ Phys. \issue(#1,#2,#3)}
\newc{\PRDR}[3]{{Phys. Rev. D} {\bf #1}, Rapid  Communications, #2 (#3)}

\def\PLB(#1,#2,#3){Phys.\ Lett.\ B  \iss(#1,#2,#3)}
\def\JHEP(#1,#2,#3){JHEP \iss(#1,#2,#3)}

\def\amususy{a_\mu^{\rm SUSY}}
\def\gmin2{(g-2)_\mu}

\catcode`\@=11 

\def\vev#1{\left\langle #1\right\rangle}
\def\lsim{\mathrel{\mathpalette\@versim<}}
\def\gsim{\mathrel{\mathpalette\@versim>}}
\def\@versim#1#2{\vcenter{\offinterlineskip
    \ialign{$\m@th#1\hfil##\hfil$\crcr#2\crcr\sim\crcr } }}
\def\etal{{\em et. al.}}

\def\r2{\sqrt 2}
\def\beq{\begin{equation}}
\def\eeq{\end{equation}}
\def\beqn{\begin{eqnarray}}
\def\eeqn{\end{eqnarray}}
\def\sinW2{\sin^2\theta_W}

\def\mz2{M_{z}^2}
\def\c2b{\cos 2\beta}

\def\m#1{{\tilde m}_#1}

\def\mw#1{{\tilde m}_{\omega #1}}

\def\mz{M_Z}
\def\m0{m_0}
\def\mhalf{m_{\frac{1}{2}}}

\def\cb{\cos\beta}

\def\sec2w{sec^2\theta_W}

\def\amususy{a_\mu^{\rm SUSY}}
\def\gmin2{(g-2)_\mu}

\catcode`\@=11 

\def\vev#1{\left\langle #1\right\rangle}
\def\lsim{\mathrel{\mathpalette\@versim<}}
\def\gsim{\mathrel{\mathpalette\@versim>}}
\def\@versim#1#2{\vcenter{\offinterlineskip
    \ialign{$\m@th#1\hfil##\hfil$\crcr#2\crcr\sim\crcr } }}
\def\etal{{\em et. al.}}

\def\tb{\tilde b}

\def\tL{\tilde L}

\def \chonep{{\wt\chi_1}^{+}}
\def \chonem{{\wt\chi_1^-}}
\def \chonep2{{\wt\chi_2^+}}
\def \chonem2{{\wt\chi_2^-}}

\def \chonepm{{\wt\chi_1}^{\pm}}

\def \mchonepm{m_{\chonepm}}

\def \chtwopm{{\wt\chi_2}^{\pm}}

\def \lstop{\wt{t}_{1}}

\def \hstop{\wt{t}_{2}}

\def \mlstop{m_{\lstop}}
\def \mhstop{m_{\hstop}}

\def \lspone{\wt\chi_1^0}
\def \mlspone{m_{\lspone}}
\def \lsptwo{\wt\chi_2^0}
\def \mlsptwo{m_{\lsptwo}}
\def \lspthree{\wt\chi_3^0}
\def \mlspthree{m_{\lspthree}}
\def \lspfour{\wt\chi_4^0}

\def\PL{Phys. Lett.}
\def\PRL{Phys. Rev. Lett.}

\def\PR{Phys. Rev.}

\def \lsptwo{\wt\chi_2^0}
\def \lspone{\wt\chi_1^0}
\def \chonem {{\wt\chi_1^\pm}}
\def \chargino1 {{\wt\chi_1^\pm}}
\def \chargino2 {{\wt\chi_2^\pm}}
\def \lstop{\wt{t}_{1}}
\def \ch2m {{\wt\chi_2^-}}

\def \chonep {{\wt\chi_1^+}}

\hypersetup{
   bookmarks=true,         
   unicode=false,          
   pdftoolbar=true,        
   pdfmenubar=true,        
   pdffitwindow=false,     
   pdfstartview={FitH},    
   pdftitle={My title},    
   pdfauthor={Author},     
   pdfsubject={Subject},   
   pdfcreator={Creator},   
   pdfproducer={Producer}, 
   pdfkeywords={keyword1} {key2} {key3}, 
   pdfnewwindow=true,      
   colorlinks=true,       
   linkcolor=blue,        
   citecolor=red,         
   filecolor=magenta,      
   urlcolor=cyan,           
   linktocpage = true,
   }

 \def\mygraph#1#2{ \subfigure[]{
    \label{#1}
    \hspace*{-0.6in}
    \begin{minipage}[b]{0.5\textwidth}
    \centering
    \hspace*{4ex}
    \includegraphics[width=\textwidth]{#2}
    \vspace*{-4ex}
    \end{minipage}}
    \vspace*{-1ex}}
 
\begin{document}
 \vspace*{\fill}
 \vspace{-1.5in}
 \begin{flushright}
 {\tt HRI-RECAPP-2015-014}
 \end{flushright}
 \begin{center}
 {\Large \bf 
Reduced LHC constraints for higgsino-like heavier electroweakinos
}

 \vskip 0.3cm
 Manimala Chakraborti$^{a 1},$~
 Utpal Chattopadhyay$^{a 2},$~
 Arghya Choudhury$^{b 3},$\\
 Amitava Datta$^{c 4}$ and~
 Sujoy Poddar$^{d 5}$
 \vskip 0.3cm
 {\small $^a$
 Department of Theoretical Physics, Indian Association
 for the Cultivation of Science,\\
 2A \& B Raja S.C. Mullick Road, Jadavpur,
 Kolkata 700 032, India}\\
 {\small $^b$
  Regional Centre for Accelerator-based Particle Physics, \\
 Harish-Chandra Research Institute, Allahabad 211019, India
 }\\
 {\small $^c$
 Department of Physics, University of Calcutta, 92 A.P.C. Road, 
 Kolkata 700 009, India
 }\\
 {\small $^d$
 Department of Physics, Netaji Nagar Day College, Kolkata 700092, India}
 \end{center}

 \begin{abstract}
As a sequel to our earlier work on wino-dominated 
$\tilde \chi_1^{\pm}$ and $\tilde \chi_2^{0}$ (wino models), we focus on
the pMSSM models where $\tilde \chi_1^{\pm}$, $\tilde \chi_2^{0}$ and 
$\tilde \chi_3^{0}$ are either higgsino dominated (higgsino models) 
or admixtures of significant amount of higgsino and wino components 
(mixed models), with or without light sleptons. 
The LHC constraints in the trilepton channel are significantly 
weaker even in the presence of light sleptons, 
especially in the higgsino models, compared to those mostly 
studied by the LHC collaborations with wino-dominated 
$\tilde \chi_1^{\pm}$ and $\tilde \chi_2^{0}$.
The modes $\tilde \chi_3^{0}, \tilde \chi_2^{0} \rightarrow 
\tilde \chi_1^{0}~h$ with large branching ratios (BRs) are more common 
in the higgsino models and may produce spectacular 
signal in the LHC Run-II. In a variety of higgsino and mixed models
we have delineated the allowed parameter space due to the LHC constraints, 
the observed Dark Matter (DM) relic density of the universe,
which gets contributions from many novel DM producing mechanisms
i.e., the annihilation/coannihilation processes that lead to the correct 
range of relic density, and the precise measurement of the anomalous 
magnetic moment of the muon. In the higgsino models many new DM producing 
mechanisms, which are not allowed in the wino models, open up. 
We have also explored the prospects of direct and indirect detection of DM
in the context of the LUX and IceCube experiments respectively.
In an extended model having only light gluinos in addition to the 
electroweak sparticles, the gluinos decay into final states with  
multiple taggable b-jets with very large BRs. As a consequence,
the existing ATLAS data in the $0l$ + jets (3b) + \mbox{$E\!\!\!\!/_T$} 
channel provide the best limit on $m_{\tilde g}$ ($\approx$ 1.3 TeV). 
Several novel signatures of higgsino models for LHC Run-II 
and ILC have been identified.          

\end{abstract}
 \vskip 0.05cm
\hrule
 \vskip 0.05cm
\noindent
{\small $^1$tpmc@iacs.res.in, $^2$tpuc@iacs.res.in, $^3$arghyachoudhury@hri.res.in, $^4$adatta@iiserkol.ac.in,\\
$^5$sujoy.phy@gmail.com }
\newpage
\setcounter{footnote}{0}

\hrule
\tableofcontents
\vspace{0.2cm}
\hrule

\section{Introduction}
\label{Sec:Introduction}

The first phase of the p-p collision at the Large Hadron 
Collider (LHC) (Run-I) has given lower limits on the masses 
of the super-particles (sparticles) 
for models involving supersymmetry\cite{SUSYreviews1,SUSYreviews2,SUSYbooks}, 
although the latter is yet to be discovered.  
The bounds on the masses of the strongly 
interacting sparticles are already stringent
\cite{atlaspap1,atlaspap2,atlaspap3,atlaspap4,atlaspap5,
cmsall}\footnote{However in 
compressed SUSY type scenarios limits on sparticle masses are 
considerably weaker\cite{compressed}.}. 
Searches have just been started at the new round of 
experiments at higher energies (13 TeV).
However, the possibility that these 
sparticles are even beyond the reach of the ongoing 
experiments as well is wide open. It is known that this scenario 
is indeed favoured by the SUSY flavour and SUSY CP 
problems\cite{SUSYreviews2}. On the other hand, we also note that 
heavy squarks belonging to the first two generations do not 
spoil the naturalness\cite{naturalness,naturalness_recent} of a SUSY model.

It is remarkable that the observed mass of the Higgs boson at around 
125 GeV\cite{HiggsDiscoveryJuly2012} 
at CERN is well within the MSSM predicted upper 
limit of $M_h$ $(\lsim 135~{\rm GeV})$, where $M_h$ refers to the mass of the  
CP even neutral lighter Higgs boson $h$. In this analysis 
we consider the decoupled Higgs scenario of the
MSSM, namely $M_A >>M_Z, M_h$\cite{SUSYreviews2,SUSYbooks},  
where $M_A$ refers to the mass of the pseudoscalar Higgs boson. 
In the decoupling limit 
$h$ becomes Standard Model like in its couplings\cite{djouadimssmrev}.
It is well known that the current Higgs data are indeed consistent with 
the decoupling limit\cite{Flechl}.
   
If the heavy squark-gluino scenario along with a decoupled 
Higgs sector is indeed realized in nature,  
we must accept that the observability of SUSY signals hinges on 
the properties of the sparticles in the electroweak (EW) 
sector\footnote{The fermionic members of this sector, the charginos and the 
neutralinos, are referred to as the electroweakinos while the scalar members 
are sleptons of both L and R types and sneutrinos.}. 
Although the production cross-sections of 
these sparticles are rather modest, significant bounds on their masses have 
already been obtained by both the ATLAS and CMS collaborations 
of the LHC \cite{atlasew1,atlasew2,atlasew3,atlasew4,cmsew1,cmsew2}
from the null results of 
(i) chargino ($\chonepm$) - neutralino ($\lsptwo$) searches\footnote{Throughout 
this paper chargino would stand for the lighter chargino ($\chonepm$) unless otherwise 
mentioned and the four neutralinos $\lspone - \lspfour$ are arranged in 
order of ascending masses.} via the  process  
$p p \rightarrow \chonepm \lsptwo$ leading to the trilepton + transverse missing energy 
($\met$) signal and (ii) slepton searches via the opposite sign same 
flavour dilepton + $\met$ channel. We shall focus on the analyses 
performed by the  ATLAS group. They obtained model independent upper bounds on the 
cross-sections of 
these processes applicable to any Beyond the Standard Model (BSM) scenario
 corresponding to different signal 
regions  each characterized by an appropriate set of selection criteria.
The results were interpreted in terms of 
several simplified models. 
A large number of phenomenological analyses have addressed 
the electroweakino searches and related topics in the 
context of the LHC \cite{spoiler,electroweakino,older}.

The discovery potentials of the charginos and the 
neutralinos depend on their pair production 
cross-sections and  decay branching ratios (BRs) into leptonic channels which contribute to the trilepton 
signal\footnote{In this work lepton usually implies electrons and muons unless mentioned otherwise.}. 
These observables depend - among other things - on their compositions. The focus of this paper is 
on the phenomenology of models where $\chonepm$, $\lsptwo$ and $\lspthree$ are either 
higgsino dominated or admixtures of significant amount of higgsino and wino components. 
In the following, we shall refer to :\\ 
i) the former class of models as the $\chonepm${\it -higgsino} 
or simply, the {\it higgsino models}, and \\
ii) the latter class of models as  $\chonepm${\it -mixed} or simply, 
the {\it mixed models}. \\

On the other hand, the LHC collaborations 
restricted their analyses 
of chargino-neutralino search in the trilepton channel
to simplified scenarios where the 
lighter chargino ($\chonepm$) and the second lightest 
neutralino ($\lsptwo$) are wino dominated and are nearly mass degenerate. 
All models belonging to this class will be referred to as the
$\chonepm${\it -wino}
or simply, the {\it wino models}. As in the analyses of the ATLAS or the CMS 
collaborations the  lightest supersymmetric particle (LSP), i.e., 
the lightest neutralino ($\lspone$) is considered to be strongly 
bino-dominated over the parameter space with $\mchonepm >> \mlspone$. 
In the {\it higgsino models} this is ensured by the chosen 
hierarchy $\mu = M_2/2 >> M_1$, where $M_1, M_2$ and $\mu$ are the U(1),  
SU(2) gaugino and higgsino mass parameters respectively. In some regions 
where $\mu$ satisfies $\mu = M_2/2 \gsim M_1$ the
LSP gets non-negligible higgsino components. 

In the backdrop of the basic varieties for the compositions of the 
electroweakinos and 
the correlation of slepton masses with that of the electroweakinos,
that we are going to enumerate shortly,
one of the main goals of this paper is to go beyond 
the {\it wino} models and
reinterpret the ATLAS data in several {\it higgsino} and {\it mixed models}. 
Before we move on to the above models, we would like to set the background 
briefly by digging into the analysis of the $\chonepm$- {\it wino} scenario.

{\underline {\bf $\boldsymbol\chonepm$- {$\boldsymbol {wino}$}:}} 
Each simplified {\it wino model} considered by ATLAS and CMS analyses 
\cite{atlasew1,atlasew4,cmsew1,cmsew2} belongs to either of the two following 
broad categories. In one case, sleptons of all three flavours are heavier 
than the winos (the {\bf L}ight {\bf W}ino and {\bf H}eavy {\bf S}lepton model
(LWHS))\footnote{In Reference~\cite{older} this model 
was called the {\bf L}ight {\bf G}augino and {\bf H}eavy {\bf S}lepton 
(LGHS) model but we 
find the terminology LWHS to be better suited for discriminating between the {\it wino}
and the {\it higgsino models}. A similar change in nomenclature applies 
to all the models discussed in Ref.~\cite{older}.}. 
In the other category, at least one type of slepton of all 
flavours (i.e., either L  or R-type or both) is lighter than $\chonepm$ and 
$\lsptwo$. The BRs of the electroweakinos and consequently, the discovery 
channels 
in the two scenarios may be significantly different. The second category in turn 
consists of several subcategories depending on the type of the light sleptons, 
and their masses with respect to $\mlspone$ and $\mchonepm$,
some of which were considered by the ATLAS group while the CMS collaboration 
as well as the analysis of  Ref.~\cite{older} 
studied more variations. The important 
features of each subcategory will be summarized in the latter sections. 
In the {\it wino models} $\chonepm$ and $\lsptwo$ production followed by 
their decays into 
trileptons is the main discovery channel in most cases. 

We wish to stress that in Ref.\cite{older} as well as in this work 
we do not restrict ourselves to stay within 
the decoupled slepton scenario. The presence of relatively light sleptons has two important implications. First, the constraints
from the direct slepton searches at the LHC must be included in our analysis. Moreover, the possibility that the sleptons play 
active roles in DM production\footnote{{\it The DM producing mechanism} 
which would often be quoted in this work would 
mean annihilation/coannihilation processes that bring the DM 
relic density within the acceptable range given by the WMAP/PLANCK data.}
is resurrected, as has already been noted in Ref.\cite{older} and will be further illustrated in this analysis.

The slepton search results, mainly for the selectron ($\wt{e}$) and the 
smuon ($\wt{\mu}$) are fairly insensitive to the electroweakino sector. 
We particularly note that sleptons of the first two generations have negligible 
L-R mixing, which in turn means that there is hardly any dependence of $\mu$
or $\tan\beta$ in determining the masses of these sleptons or their couplings 
with the gauge bosons.  Any slepton lighter than $\chonepm$ or $\lsptwo$ 
decays into its fermionic superpartner and 
$\lspone$ with 100\% BR and this is independent of the composition
of $\chonepm / \lspone$. Subject to these general 
assumptions mass bounds were obtained by the LHC collaborations 
in several simplified models\cite{atlasew2,cmsew2}.
In view of the above independence 
we shall directly use 
the constraints from the slepton searches as derived in Ref.~\cite{older} 
for the {\it wino models}.

The  models studied by the ATLAS collaboration\footnote{Similar simplified models were also analysed 
by the CMS collaboration.} to interpret their search results are in some sense oversimplified. The  
parameters of such a  model provide a minimal set to understand important aspects of a  SUSY signal. 
However, in order to test SUSY in the light of the LHC as well as other constraints from the so called 
indirect tests,  one requires a  closely related but a complete model 
like the phenomenological minimal supersymmetric standard model 
(pMSSM)~\cite{pmssm} with additional parameters. 
In Ref.~\cite{older} we enlarged each simplified {\it wino 
model} by introducing a minimal set of parameters belonging to the EW sector 
(like $\mu$ and $\tan\beta$) and tried to scrutinize the pMSSM thus obtained
by considering three major constraints, namely, the LHC mass limits on the 
chargino-neutralino and the slepton sectors, measured 
dark matter (DM)\cite{dm_rev1,Silk,dm_rev2} 
relic density of the universe from the WMAP\cite{wmap}/PLANCK\cite{planck}\footnote{A partial 
list of works on SUSY DM may be seen in 
Ref.~\cite{dmmany,dmsugra,dmmssm,dmmssmfurther}.} and the precisely measured 
value of $\gmin2$ \cite{g-2exp}. 
Moreover, we dispensed with the unrealistic assumption 
$m_{\snu} = m_{\slepton_L }$ which leads to erroneous LHC limits especially 
if $\mlsptwo \approx m_{\slepton_L }$. These  modifications change the
LHC limits quite significantly in some models and 
we computed these changes in Ref.~\cite{older} by a PYTHIA (v6.428)\cite{pythia} based 
analysis using ATLAS data. New bounds for 
several wino-slepton mass hierarchies not considered by 
the ATLAS collaboration were also derived. For each model,
compatibility with the three major constraints delineates an allowed 
parameter space (APS). 
Each APS, in turn, enables us to focus on the expected SUSY 
signals in the future LHC experiments. 

{\underline {$\boldsymbol \chonepm$- {$\boldsymbol {higgsino~or~mixed}$}:}} 
In this paper we extend our earlier analysis to
the  $\chonepm${\it -higgsino} and  $\chonepm${\it -mixed models}. 
The mass hierarchies among the sleptons and electroweakinos are, 
however, similar to the ones in \cite{older}. 
In Table~\ref{summarytab1}
we present the models analyzed in this paper and the choice of parameters for each of them. In our analysis, $M_1$ and $M_2$ are free parameters.
\begin{table}[!htb]
\begin{center}
\begin{tabular}{|c||c|c|c|}
\hline
   & Model & Acronym & Parameter Choice \\
\hline
\hline
\multirow{6}{*}{$\chonepm$-Higgsino} &  Light Higgsino and light & LHLRS & $\mu = \frac{M_2}{2}$        \\
                                     &  Left and Right Sleptons  & (Sec.\ref{sec:LHLRS})      
& $m_{\wt l_L} = m_{\wt l_R} = (\mlspone + \mchonepm)/2$   \\
\cline{2-4}
                                     &  Light Higgsino and light & LHLS & $\mu = \frac{M_2}{2}$        \\  
                                     &  Left Sleptons            & (Sec.\ref{sec:LHLS})
& $m_{\wt l_L} = (\mlspone + \mchonepm)/2, m_{\wt l_R} =$ 2 TeV \\
\cline{2-4}
                                     &  Light Higgsino and Heavy & LHHS & $\mu = \frac{M_2}{2}$  \\
                                     &  Sleptons                 & (Sec.\ref{sec:LHHS}) 
& $m_{\wt l_{L,R}}= \mu +$ 200 GeV \\
\hline
\multirow{4}{*}{$\chonepm$-Mixed}   &   Light Mixed and light & LMLRS  & $\mu = 1.05 M_2$ \\
                                    &   Left and Right Sleptons   & (Sec.\ref{sec:LMLRS})      
& $m_{\wt l_L} = m_{\wt l_R} = (\mlspone + \mchonepm)/2$ \\
\cline{2-4}
                                    &   Light Mixed and light & LMLS  & $\mu = 1.05 M_2$ \\
                                    &   Left Sleptons      &  (Sec.\ref{sec:LMLS})     &
$m_{\wt l_L} = (\mlspone + \mchonepm)/2, m_{\wt l_R} =$ 2 TeV  \\
\hline
\end{tabular}
\end{center}
\caption[]{Summaries of the models analyzed in this work. The parameter 
choice for each case is presented in the last column. For all the analyses we 
take  
$M_1 << M_2$, $M_1 \lsim \mu $ to make the LSP predominantly a bino. Two
representative values of $\tan\beta = $ 6 and 30 are considered in this analysis.
For the LHHS model, however, we consider only $\tan\beta =$ 30 case for reasons discussed
in the text.}
\label{summarytab1}
\end{table}

As mentioned earlier
the LSP is also assumed to be bino-dominated with some degree of higgsino mixing, 
depending on the parameter space.
Finally, we delineate the APSs in both the  models while isolating 
the effects of each major constraint clearly. 
We emphasize that in the post-LHC era, these  models, especially 
the {\it higgsino models}, have not received due attention in the literature. 
Yet, the difference in phenomenology 
of the {\it higgsino models} and that of the {\it wino models} is indeed worth 
noting. Firstly, the LHC exclusion contours from the 
trilepton searches shrink  significantly in the 
{\it higgsino models} and even become irrelevant in some scenarios. 
As we will see later 
the physics of DM relic density and $\gmin2$  in the {\it higgsino models} 
are also quite distinctive. 
All these points will be elaborated in the rest of this paper.

We will further confront each APS thus obtained with other constraints like those from 
direct\cite{xenon100,LUX} 
and indirect\cite{Icecube1,Icecube2} dark matter searches. 
We include current limits as well as 
compare our results in relation to future reaches of these 
experiments\cite{xenon1t,Icecube1}. We also keep in mind the sizable 
uncertainties involved in these constraints 
(see Ref.~\cite{older} and the references therein).
We emphasize that unlike the analysis of Ref.~\cite{older}, here we have a sizable amount of higgsino
content within the LSP because $\mu$ and $M_1$ are not widely
separated. This leads to a considerable increase in the 
spin-independent LSP-nucleon
scattering cross-section. Additionally, a larger higgsino
content within the LSP generically 
increases the spin-dependent LSP-nucleon cross-section. This,
in turn, increases the gravitational capture cross-section of the LSPs
within astrophysically dense regions like the core of the Sun. In addition,  
the LSPs with larger
higgsino content may potentially contribute to neutrinos created within the
Sun via LSP pair annihilation. Thus, the IceCube experiment puts limits on 
the muon flux at the detector site which are not far away from the predictions of the models.

We next consider an extended scenario in which only one strongly interacting sparticle is 
within the reach of the LHC-13/14 TeV experiments and assume this sparticle to be the gluino. 
The purpose is to study the feasibility of characterizing different {\it higgsino models} 
from their gluino decay signatures. We explore the gluino mass limits obtained at the 
LHC-8 TeV experiments via the $n$-leptons + $m$-jets (with or without b-tagging) + $\met$ signals with 
different values of $m$ and $n$\cite{atlaspap1,atlaspap2,atlaspap3,atlaspap4}. 
By selecting a few benchmark points (BPs) with different characteristics
we compute the revised gluino mass limits at the generator level by using PYTHIA (v6.428)\cite{pythia}. 
This gives the sensitivity of various {\it higgsino models} to signals with different  
values of $m$ and $n$  and helps to anticipate the future search prospects. 

The plan of the paper is as follows. We note that because of 
the enhancement of higgsino components within
$\lsptwo$, $\lspthree$ and $\chonepm$, the  
production cross-sections of $\chonepm$-$\lsptwo$ and $\chonepm$-$\lspthree$
in the $\chonepm${\it -higgsino ($\chonepm${\it -mixed}) models} are 
significantly (modestly) reduced from those in the $\chonepm${\it -wino models}. 
The characteristics of various cross-sections will be discussed in 
Sec.~\ref{Sec:ProductionElectroweakinos} with numerical examples. 
The relevant constraints for the {\it higgsino} and {\it mixed} models
as well as the procedures for parameter space scanning and simulation of the trilepton signal
will be discussed in Sec.\ref{Sec:Methodology}.
The parameter spaces allowed by the main constraints in the 
$\chonepm${\it -higgsino} and $\chonepm${\it -mixed models}
characterized by different mass hierarchies among the sleptons and the elctroweakinos will
be presented in Secs. \ref{Sec:HiggsinoAndMixedLRS}-\ref{sec:LHHS}.
We will explore the prospects of direct and indirect detection of dark matter
in Sec.~\ref{Sec:DirectDMSIDetection} and Sec.~\ref{Sec:DirectDMSDandIndirect}
respectively. The extended models with a light gluino
and electroweak sparticles, introduced in the last paragraph, will be taken up in 
Sec.~\ref{Sec:ConstraintsGluino}. Our main results and the conclusions are summarized in 
Sec.~\ref{Sec:Conclusion}.

\section{Production of chargino-neutralino pairs in different models}
\label{Sec:ProductionElectroweakinos}
The size of the  chargino - neutralino production cross-section with 
different neutralinos accompanying $\chonepm$ is significantly different in 
the {\it wino}, {\it higgsino} and {\it mixed models}. In the {\it wino models} 
the main signal, namely $3l + \met$ comes from $p p \rightarrow \chonepm \lsptwo$,
since the production of the heavier neutralinos like $\lspthree$ or $\lspfour$  
is highly suppressed. It may be recalled that $\chonepm$ and $\lsptwo$ masses 
are  degenerate in this case and are controlled by the soft breaking mass 
$M_2$ for the SU(2) gauginos while  the masses of $\lspthree$ and $\lspfour$ 
are governed by the higgsino mass parameter $\mu$ with $\mu >> M_2$. This 
degeneracy holds for all $\mchonepm$ to a very good approximation.

\begin{table}[!htb]
\begin{center}
\small
\begin{tabular}{|c||c|c|c||c|c|c||c|c|c|}
\hline
Masses	 & \multicolumn{3}{c||}{P1} & \multicolumn{3}{c||}{P2} & \multicolumn{3}{c|}{P3}\\
\cline{2-10}
and & \multicolumn{3}{c||}{Model} & \multicolumn{3}{c||}{Model}  & \multicolumn{3}{c|}{Model}  \\
\cline{2-4}
\cline{5-7}
\cline{8-10}
cross-sections     &  Wino   & Mixed  & Higgsino & Wino   & Mixed  & Higgsino &  Wino   & Mixed  & Higgsino \\
\hline
$\mlspone$ 	& 150  & 150    & 150   &150   &150  &150 &150   &150  &150 \\
\hline
$\mchonepm$ & 200   & 200  & 200   &300   &300  &300  & 650 & 650 & 650 \\
\hline
$\mlsptwo$ 	& 201    &210   & 219  &300 &304  &304 &650  &651  &652 \\
\hline
$\mlspthree$ &421 & 269  & 221  & 604   &370  &312  &1256  &722  &657 \\
\hline
$\sigma_(pp \ra \chonepm \lsptwo)$ &0.7  &0.43   &0.194  &0.129   &0.083  &0.037 &0.00207  &0.00135 &6.9$\times 10^{-4}$ \\
\hline
$\sigma_(pp \ra \chonepm \lspthree)$& $10^{-3}$ & 0.06 & 0.209 &$10^{-4}$ &0.011 &0.037 & -  &0.00022& 6.63$\times 10^{-4}$   \\
\hline
$\sigma_{Total} $   &0.70  &0.49  &0.403  &0.129  &0.094  &0.074 &0.00207   &0.00157 &0.00135 \\
\hline
       \end{tabular}
       \end{center}
           \caption{Table showing the relevant masses and the cross-sections 
for three parameter points P1, P2 and P3. Here all the masses are in GeV and cross-sections are in pb. }
\label{cstable}
          \end{table}

In contrast, the  {\it higgsino models} are characterized by $\mu << M_2$. 
As a result, $\lsptwo$ and $\lspthree$ are nearly mass-degenerate with $\chonepm$ 
where all masses are essentially determined by $\mu$. The degeneracy is more 
exact as $\mchonepm$ increases. Here both  $\chonepm \lsptwo$ and  $\chonepm \lspthree$ 
production cross-sections are significant. In spite of this, the total chargino - neutralino 
production cross-section is smaller than that in a generic {\it wino model} with 
similar chargino and neutralino masses. In the {\it mixed models} ($\mu \approx M_2$),  
the cross-sections  typically have intermediate values with respect to those of the
{\it wino} and the {\it higgsino models}. The $\chonepm$ and $\lsptwo$ 
are nearly degenerate and the degree of degeneracy increases as $\mchonepm$ increases. 
There is always a much larger mass difference between $\lsptwo$ and $\lspthree$ 
compared to the {\it higgsino model}. It may be noted that these features are  
independent of $M_1$, the $U(1)$ gaugino mass, as long as the LSP is bino-dominated, 
i.e, $M_1 << M_2, \mu$. The choice of $\tan\beta$, the ratio of the Higgs 
vacuum expectation values, only has a marginal impact. 

Production of $\chonepm-\lsptwo/\lspthree$ occurs through the
process $q_i \bar q^\prime_i / q^\prime_i \bar q_i \rightarrow \chonepm \wt \chi_{2,3}^0$,
where $q$ and $q^\prime$ for the first
two generations ($i = 1, 2$) refer to up and down type of quarks respectively.
When the first two generations of squarks are heavy, s-channel W boson exchange becomes
the dominant production process.
We note that the $W^{\pm}-\chonepm-\wt \chi_{2,3}^0$ coupling is contributed by terms 
involving products of wino components as well as terms having products of 
 higgsino components of the relevant electroweakinos. The former terms typically 
dominate over the latter. Thus, as we move from the
{\it wino model} to the {\it higgsino model}, the gradually
diminishing wino contents of $\chonepm$ and $\wt \chi_{2,3}^0$ render smaller and
smaller cross-sections for $\chonepm-\lsptwo/\lspthree$ production.

The above features are illustrated by Table \ref{cstable}. Here the LSP 
mass is fixed at 150 GeV. The parameters $M_2$ and $\mu$ are differently 
chosen in the three models, in particular, $\mu = 2 M_2$ (wino), $\mu = M_2 / 2$ ({\it higgsino}) 
and $\mu \simeq M_2$ ({\it mixed}). Throughout this paper we shall present the 
numerical results with these characteristic choices. It follows from Table \ref{cstable} 
that for a wide range of the chargino mass the total chargino - neutralino 
production cross-section  in the {\it higgsino (mixed) model} is 60-65\% (70-75\%) 
of that in the {\it wino model}. All next to leading order (NLO) cross-sections are calculated with 
PROSPINO 2.1~\cite{prospino}. The reduction in the production cross-section 
is one of the reasons for relaxed mass limits in the {\it higgsino} and 
{\it mixed models}. However, the LHC mass limits are also sensitive to the choice of the 
slepton sector which will be addressed in Secs.~\ref{Sec:HiggsinoAndMixedLRS},
\ref{Sec:HiggsinoAndMixedLS} and \ref{sec:LHHS}.
\section{The Methodology}
\label{Sec:Methodology}
In this section we summarize the constraints that we use 
to restrict  the  parameter spaces of several {\it higgsino}
and {\it mixed models}. We also present
 brief sketches  of the simulation using PYTHIA (v6.428)
 as well as the procedure for scanning the parameter space. 
\subsection{The Constraints}
The three entries listed below are characterized by 
relatively small theoretical/experimental uncertainties and they 
constitute what we call the three major constraints.
\begin{enumerate}
\item The LHC constraints from the chargino-neutralino searches in the trilepton channel used in this paper are from 
the ATLAS conference report\cite{atlas3lew} with 
$\lum=$ 20 $\ifb$ data, which is the source of the published paper\cite{atlasew1}. Similarly, for 
the constraints in the slepton sector 
we use the conference report\cite{atlas2lew}. It may be noted that there is no major difference in the exclusion
contours among the published versions and their predecessors. The ATLAS collaboration quotes
the upper limits on the number of events in any new physics model at 95\% CL subject to  different sets of selection criteria 
(see the next subsection). We have simulated the trilepton signal using the same sets of kinematical selections
at the generator level using PYTHIA (v6.428) for  the 
{\it higgsino} and {\it mixed models} analysed in this paper. 
The model independent limits then enable us to sketch  the exclusion contours in the $\mchonepm - \mlspone$
plane in different models for representative choices of the other parameters as 
detailed in the following sections. As already noted in the 
introduction the slepton constraints hardly depend on the composition of the charginos and neutralinos heavier 
than the sleptons. We have, therefore, directly used the  exclusion contours 
from light L and LR slepton searches in the corresponding {\it wino models} obtained in \cite{older}.
\item The anomalous magnetic moment of the muon ($a_\mu=\frac{1}{2}\gmin2$)\cite{muonrev}
is an important probe for new physics beyond the standard model (SM). 
There is a significantly large deviation
(more than 3$\sigma$) of the SM prediction\cite{g-2sm1,g-2sm2} from the experimental data\cite{g-2exp}.
Contributions to $a_\mu^{\rm SM}$ can be categorized into three parts : a part coming from
pure quantum electrodynamics, electroweak contributions and a hadronic part.
SUSY contributions to $a_\mu$, namely $\amususy$, scale with $\tan\beta$.
It can also be large when 
chargino, sneutrino, neutralino and smuons are light\cite{oldSusyMuong}.  
Thus, it is possible to constrain the SUSY parameter space effectively with given upper and lower limits of
$\Delta a_\mu=a_\mu^{\rm exp}- a_\mu^{\rm SM}$. The deviation of the experimental 
data from the SM calculation amounts to\cite{g-2sm2} 
\begin{equation}
\Delta a_\mu=a_\mu^{\rm exp}- a_\mu^{\rm SM}=(29.3 \pm 9.0)\times 10^{-10}.
\end{equation}
A partial list of analyses regarding $\amususy$ in SUSY models is provided by
 Refs. \cite{oldSusyMuong,susyg-2A,susyg-2B,endo}.  
With the Higgs mass at 125 GeV and stringent lower bounds on squark-gluino masses 
coming from the LHC, simplified models like mSUGRA have become rather inefficient to 
accommodate the $\gmin2$ anomaly\cite{baer-prannath}.
However, non-universal models can still successfully explain the above range of 
$\Delta a_\mu$\cite{Nonunivg-2}. It should be noted that the $\gmin2$ constraint is able to impose
definite upper and lower bounds on the sparticle masses~\cite{g-2sparticlebounds}.
\item Following Ref.\cite{older}, relic density limits from the WMAP/PLANCK \cite{wmap,planck} is taken as,
\begin{equation}
0.092 < \Omega_{\tilde \chi} h^2 <0.138.
\label{planckdata}
\end{equation}
\end{enumerate}
Apart from the above direct constraints we will further analyze our 
results in relation to the following dark matter detection limits.
\bi
\item The direct detection bound on spin-independent (SI) LSP-proton scattering cross-section
$\sigma_{\tilde \chi p}^{\rm SI}$ is imposed using the LUX\cite{LUX} data.
\item  DM indirect detection constraints like the bounds on 
spin-dependent (SD) 
LSP-proton scattering cross-section $\sigma_{\tilde \chi p}^{\rm SD}$ and limits on muon flux 
given by the IceCube\cite{Icecube1} are also important in our case.
\ei
We use SuSpect version 2.41~\cite{suspect} for spectra generation and $\amususy$ calculation.
SUSYHIT~\cite{susyhit} is used for obtaining the decay BRs of the sparticles. DM 
relic density and observables related to its direct and indirect detection are 
computed using micrOMEGAs version 3.2 \cite{micromega3}.
\subsection{The Simulation}
In Ref.\cite{atlas3lew} the ATLAS collaboration defined six signal regions (SRs) : 
SRnoZa, SRnoZb, SRnoZc, SRZa, SRZb and SRZc. 
Table 1 of Ref.\cite{atlas3lew} includes the details of the cuts corresponding to each signal region.
The corresponding upper limits on the number of new physics events are listed in Table 4 of Ref.\cite{atlas3lew}.
In all models analysed in this paper we
simulate the trilepton signal for the above SRs with a given  $\mchonepm$ by increasing the LSP mass in small steps.
Below a certain LSP mass the above ATLAS upperbounds on the number of new physics events are violated for at least one SR.
A point on the exclusion contour is determined in this way. On the other hand when $\mchonepm$ is varied, all the 
LSP masses are allowed
above a certain $\mchonepm$ which is the lower limit on this parameter. The validation of 
our simulation and other details have been described in Ref.\cite{older}.
We use the same setup for the present work.

\subsection{Scanning the parameter space}  
We take the strong sector parameters to be heavy by choosing $M_3 = 2$ TeV and fixing
the masses of the first two generations of squarks and the mass ($M_A$) of 
the pseudoscalar Higgs boson at 3 TeV. 
The mass parameters for the third generation of squarks  are 
fixed at 1.2 TeV. The top trilinear parameter 
$A_t$ is varied in the range -5 TeV $< A_t <$ 5 TeV in order to obtain the 
lighter Higgs scalar mass to be in the interval
$122<m_h<128$~GeV. 
All the other trilinear parameters except $A_t$ are taken to be zero.
A theoretical uncertainty of 3~GeV in the computation of Higgs mass 
is considered here.
This spread in $m_h$ arises from the uncertainties in the higher order loop corrections
up to three loops, that due to the top-quark
mass, renormalisation scheme, scale
dependence etc\cite{higgsuncertainty3GeV}.  The Higgs bosons other than the lightest one are 
assumed to be decoupled.
We perform scan over the parameters $M_1$, and $M_2$ in the range
100~GeV-1~TeV. The slepton masses are correlated to 
$\mlspone$ and $\mchonepm / \mlsptwo$
in this work and the nature of correlation are described in the appropriate 
subsections. Values of the relevant SM parameters are taken 
as $m_t^{pole}=173.2$~GeV,
$m_b^{\overline {MS}}=4.19$~GeV and $m_\tau=1.77$~GeV. Finally, 
we consider only the positive sign of $\mu$ in this analysis. 

\section{The $\chonepm$-Higgsino and $\chonepm$-Mixed Models
with Light Left and Right Sleptons}
\label{Sec:HiggsinoAndMixedLRS}
In this sequel analysis that extends our previous work \cite{older} 
on $\chonepm$-{\it wino models} towards $\chonepm$-{\it higgsino}
ones we will review a few salient points as well as refer to some figures 
of the earlier analysis for the sake of clarity
and easier understanding of the present work.
The analysis of various $\chonepm$-{\it higgsino} and $\chonepm$-{\it mixed models} 
using the methodology sketched in the last section, will be presented
in this and the next two sections.
\subsection{A brief review of the Light Wino and light Left 
and Right Sleptons (LWLRS) model}
\label{sec:LWLRS}
In the {\bf L}ight {\bf W}ino 
and light {\bf L}eft and {\bf R}ight {\bf S}leptons (LWLRS) model\cite{older}, 
the L and the R types of sleptons were assumed to be mass degenerate 
(modulo the D-term contributions) with a common mass :
($ x_1 ~M_1 + x_2 ~ M_2$). Three choices were considered: 
i) $x_1 =x_2 = 0.5$,  ii) $x_1 = 0.25, x_2 = 0.75$ and
iii) $x_1 = 0.75, x_2 = 0.25$, with the slepton mass lying between $M_1$ and $M_2$.
The models with $x_1 \neq x_2$ are referred to as tilted models denoted 
by the LWLRS$_{\chonepm}($LWLRS$_{\lspone}$) 
with the slepton mass closer to the $\chonepm$ ($\lspone$)  mass. 
The ATLAS group did not interpret their data in any form of the LWLRS model. 
However, as we shall show in the next subsection, some versions of the corresponding 
{\it higgsino model} are indeed rather intriguing since they are practically unconstrained by the LHC data.

The impact of the three major constraints (see Secs.\ref{Sec:Introduction} 
and \ref{Sec:Methodology}) 
on the parameter space of the LWLRS model for low $\tan\beta$ (= 6) 
may be seen in Figure~4a of Ref.\cite{older}.  
For a negligible LSP mass, the acceptable value of $\mchonepm$ is above 610 GeV. 
On the other hand, for higher values of the LSP masses the limits that had been 
obtained primarily from the trilepton searches earlier, 
became weaker. There were 
two branches in the parameter space consistent with the
WMAP/PLANCK data\cite{wmap,planck}. This feature is common to most of 
the {\it wino models} with low $\tan\beta$. 
The fact that relic density limit is satisfied in the upper branch 
is related to LSP-sneutrino coannihilation.
The lower branch is consistent with the relic density limits via 
LSP pair-annihilation into the Higgs resonance, corresponding to 
$\mlspone \approx m_h /2$\footnote{In some of the cases the Higgs resonance
region is accompanied by a small Z resonance annihilation branch. However, we
will mostly focus on the former in our present analysis.}. 
In spite of satisfying the DM relic density limits, a 
large part of the above parameter space corresponding to 
low $\mchonepm$ is disfavoured by the trilepton and 
slepton searches. This  indeed is the case in all the {\it wino 
models} with $x_1 = x_2 = 0.5$. 
However, the constraints from slepton searches are significantly 
relaxed in the LWLRS$_{\chonepm}$ model and one obtains $\mchonepm \geq$ 450 GeV 
for negligible LSP masses(see Fig.~6a of Ref.\cite{older}). 
The 
results of Ref.\cite{older} derived from slepton searches 
are being readily adapted in the corresponding 
{\it higgsino} or {\it mixed model} of the present analysis for reasons already discussed in 
Sec.~\ref{Sec:Introduction}. 
However, the entire parameter space allowed even by the relaxed constraints from 
slepton searches is consistent with the $\gmin2$ data at best at the 3$\sigma$ level.
This tension at low $\tan\beta$ from the $\gmin2$ constraint exists in  
all the {\it wino models} as already been noted\cite{older}. 
This feature is also shared by the {\it higgsino models} as 
we shall see in the following sections. 

A choice of a large $\tan\beta$ (= 30) in the LWLRS model 
hardly changes the LHC constraints, as may be seen 
in Fig.~4b of Ref.\cite{older}.
However, a significant amount of parameter space is discarded because 
${\tilde \tau}_1$ becomes the LSP as a result of  
enhanced degree of Left-Right mixing considering a large value of $\tan\beta$.
On the other hand, since $\stauone$ is significantly lighter 
than the other sleptons, $\tau$-rich final states deplete a 
part of the trilepton signal\footnote{The potential of final states with multiple $\tau$s 
from $\chonepm$ and $\lsptwo$ decays as new signals at the LHC was emphasized 
in Ref.\cite{Bhattacharyya:2009cc} .}.
This depletion, however, is more effective in the
parameter space beyond the reach of the LHC Run-I experiments.
We should here point out that
the small mass difference between $\tilde \tau_1$ and $\lspone$ which results into satisfying DM relic 
density via 
LSP-stau coannihilation (see 
the upper branch of the region allowed by the WMAP/PLANCK data in Fig.~4b of Ref.\cite{older}) 
makes the tau-tagged signatures difficult to observe. The Higgs resonance region 
on the other hand disappears due to a vanishingly small higgsino component 
of $\lspone$ which severely suppresses the $ h\lspone\lspone$ coupling at high 
$\tan\beta$, a generic feature shared by all the {\it wino models} with a bino-like LSP.
As expected the tension with the $\gmin2$ constraint is relaxed in this high $\tan\beta$
scenario. As a result there is a narrow  APS consistent with all the 
major constraints.

\subsection {The Light Higgsino and light Left and Right Sleptons 
(LHLRS) model}
\label{sec:LHLRS}
In this subsection we focus on the
models characterized by 
{\bf L}ight {\bf H}iggsino and light {\bf L}eft and {\bf R}ight 
{\bf S}leptons (LHLRS), where the 
sleptons are assumed to have masses nearly halfway between the masses 
of the lightest neutralino and the lighter chargino\footnote{Placement 
of slepton masses follow similar relationship with the masses of the LSP 
and the lighter chargino as mentioned in the previous subsection. This 
is also accompanied by similar tilted scenarios like the LHLRS$_{\chonepm}$ and
LHLRS$_{\lspone}$ models where sleptons are closer in masses to that of 
$\chonepm$ and $\lspone$ respectively by specific amounts.} 
while $\chonepm$ and $\lsptwo$ are higgsino-dominated in nature.
In this scenario with $M_1<\mu<M_2$,  
$\mchonepm$, $\mlsptwo$ and $\mlspthree$ are close to
$\mu$. In this paper we restrict ourselves to the regions of parameter space 
where the LSP is either bino-dominated ($M_1 << \mu$) 
or a strong admixture of 
bino and higgsinos with the former field dominating over the others. 
In the latter case, the LSP and $\chonepm$ are almost mass degenerate. 
 
We do not however include the 
scenario where the LSP is purely a higgsino or higgsino-dominated 
($\mu \lsim M_1)$. In this parameter space all the lighter electroweakinos including 
the LSP are higgsino-like and nearly mass degenerate. As a result,
no interesting LHC signature, except for the well-known monojet + 
$\met$  signal, is expected. DM searches by the LHC collaborations 
in this channel yielded only null results\cite{lhcdmsearch}.
However, the results have not been interpreted 
in terms of the pMSSM scenarios. According to the 
analysis of Ref.\cite{barducci} using 
the LHC Run-I data the bound on the LSP mass in this case is rather weak. 
Moreover, for the electroweakino
mass ranges of current phenomenological importance, the DM is found 
to be underabundant when 
$\mu \lsim M_1$\cite{higgsinodm}.  \\

\begin{figure}[!htb]
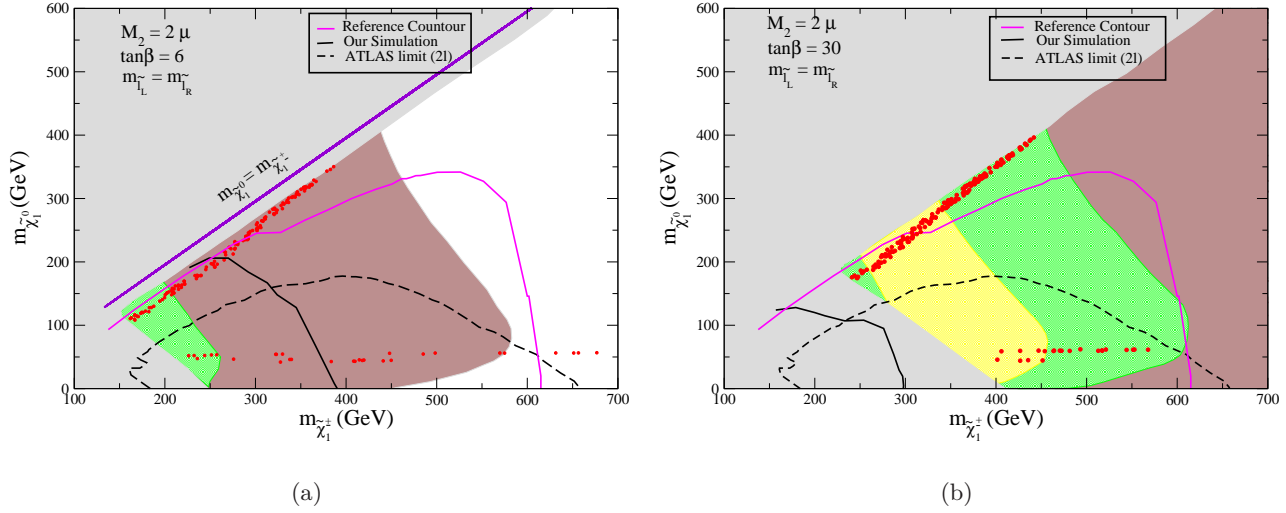

\vspace*{-0.05in}
\mygraph{MH6LR}{MH6LR.eps}
\hspace*{0.5in}
\mygraph{MH30LR}{MH30LR.eps}
\caption{The brown, green and yellow regions show the parameter space consistent
with the $\gmin2$ constraint at 3$\sigma$, 2$\sigma$ 
and 1$\sigma$ level respectively.  
The magenta line is the reference contour which represents the strongest
$\mlspone-\mchonepm$ mass limit as obtained in the 
{\bf L}ight {\bf W}ino and light {\bf L}eft {\bf S}lepton (LWLS) model 
[see Fig.~1a of Ref.\cite{older}]. 
The black line is the contour obtained by our simulation using ATLAS data. 
The black dashed line represents the exclusion contour from ATLAS slepton searches. 
The grey region to the left and above the coloured portion of the 
parameter space is either theoretically 
discarded or disallowed by our requirement of having a bino-dominated LSP 
or disallowed for the requirement of $\lspone$ to be the LSP.  
The red points satisfy the WMAP/PLANCK data on DM relic density. 
The $\mchonepm = \mlspone$ line is also shown for the plot in the left panel.
For the plot in the right panel, this line does not exist since ${\tilde \tau}_1$ 
becomes the LSP for $\mchonepm \sim \mlspone$ because of large mixing.}
\label{MH_LR}
\end{figure}

In Fig.~\ref{MH6LR}, we present the results of the LHLRS analysis
for tan$\beta =6$. 
The brown and the green regions signify the parameter space consistent
with the $\gmin2$ constraint at $3\sigma$ and $2\sigma$ 
levels respectively.  
The magenta line represents the reference contour which is the strongest
$\mlspone-\mchonepm$ mass limit as obtained in the 
{\bf L}ight {\bf W}ino and light {\bf L}eft {\bf S}lepton (LWLS) model 
[see Fig.~1a of Ref.\cite{older}]. 
The black line is the contour obtained by our simulation for the LHLRS model using ATLAS data. The 
black dashed line represents the exclusion contour from ATLAS slepton searches.
The grey region to the left of the APS is theoretically disallowed.
The small change in this region compared to that of Fig.~4a 
of Ref.\cite{older} is due to the choice of a smaller 
stop mass (here $\mlstop \simeq$~1~TeV, the same in 
the previous analysis was 2~TeV) and its effect 
on the electroweak symmetry breaking (EWSB) scale,
$Q=\sqrt{\mlstop \mhstop}$, as used in the code SuSpect\cite{suspect}. 
The grey region just below the $\mchonepm = \mlspone$ line 
corresponds to $\mu \lsim M_1$ which leads to a higgsino-dominated LSP, a 
scenario kept out of the domain of our analysis for reasons already 
discussed. The grey region above the aforesaid line is discarded 
because $\lspone$ is chosen to be the LSP, a candidate for DM.  
We should mention that the theoretically disallowed  
regions which we will often quote in this work 
arise from the facts mentioned above. 
We also take into account the LEP limits\cite{lepsusy} for sparticle 
masses, if necessary.

It is apparent that
apart from a small region of the parameter space corresponding 
to low $\mchonepm$, the collider limits are relaxed significantly with respect to
the wino dominated scenario (Fig.~4a of Ref.\cite{older}).
For negligible $\mlspone$, a substantially weaker bound ($\mchonepm \gsim 380$~GeV)
compared to the {\it wino model} is obtained via the trilepton signal. However,
the slepton mass limits discard $\mchonepm$ up to 600~GeV. 
On the other hand, with a modestly increased value of 
$\mlspone$ such as 150~GeV, the lower bound on $\mchonepm$ is about 
330~GeV in order to satisfy both 
the limits from trilepton and slepton searches.
For larger 
LSP masses, the chargino mass bounds are even weaker, as expected, and eventually disappears
for $\mlspone \gsim 200$~GeV.  
In addition to the suppression of the total chargino-neutralino
production cross-section as discussed in the Sec.~\ref{Sec:ProductionElectroweakinos}, 
the decrease in the leptonic BRs of the electroweakinos 
reduces the trilepton signal. 
\noindent
For small $\mchonepm$, the decays like  
 $\chonepm \ra W^{\pm} \lspone$ and $\lsptwo / \lspthree \ra h/Z~\lspone$ are 
kinematically forbidden. Hence, the number of two-body leptonic decays of
$\chonepm$ and $\lsptwo$ remain practically unaltered 
compared to the {\it wino model}. Moreover, the
lepton pairs from the decays of $\lspthree$ compensate the overall decrease in
the cross-section and this leads to almost unchanged LHC limits. 
The situation changes significantly when the decays $\chonepm \ra W^{\pm}~\lspone$ and $\lsptwo / \lspthree \ra h/Z \lspone$ are 
kinematically allowed. 
The BR($\chonepm \ra W^{\pm}~\lspone$) being dominant ($\simeq$ 75\%), the overall
lepton fraction drops down as $W$-boson decays with the usual  
small leptonic BRs\footnote{This happens in spite of the 
presence of  light sleptons which do not couple favourably 
to the higgsino like $\chonepm$ and $\lsptwo$.}.
Moreover, $\lsptwo$ and $\lspthree$ decay 
dominantly into the gauge/Higgs boson channels, thus
depleting the trilepton signal further\footnote{Numerical values 
of the relevant BRs for selected BPs will be provided in 
Sec. \ref{Sec:ConstraintsGluino}.}.

Regarding the constraint from $\gmin2$, the major contribution  
comes from the neutralino-smuon loop processes. However, the contribution
from the chargino-sneutrino loop diagrams are hardly ignorable. 
In this low $\tan\beta$ scenario of Fig.~\ref{MH6LR}, the $\gmin2$ constraint which typically is 
satisfied only at the level of 3$\sigma$. 
As expected, the SUSY contribution to $\gmin2$ 
falls on the lower side of the 3$\sigma$ limit rather than on the higher side.
In other words, keeping in mind the level of
uncertainties of both the SM prediction and the experimental 
data\cite{g-2sm1,g-2sm2,muonrev}, the above 3$\sigma$ zone is indeed close to the zone where agreement with the experimental data is at higher level of confidence. The situation 
would change with more accurate SM predictions and more precision 
in the experimental front.  
We simply like to comment here that accepting such 
a relaxed level of the $\gmin2$ constraint may be worthwhile if there is a scope 
to open up a region of parameter space having important LHC signature(s). 
As discussed above, in this 3$\sigma$ allowed parameter space, the 
decays $\chonepm \ra W \lspone$ and 
$\lsptwo \ra h \lspone$ would potentially lead 
to novel collider signatures like $W h \lspone$ during the LHC Run-II.

There are two separate branches 
consistent with the WMAP/PLANCK data in Fig.~\ref{MH6LR}.
In the parameter space with small  slepton masses, bulk 
annihilation\footnote{LSP-pair annihilation 
that occurs via t-channel slepton exchange.} of $\lspone$ 
may partly serve as the cause to satisfy the relic 
density limits via enhanced LSP pair-annihilation. 
But, this region of the parameter space is 
disfavoured by the LHC constraints anyway.
In the upper branch, $\mu$ is close to $M_1$ so that
the bino-like LSP has a sizable higgsino component.    
Thus, here one finds $\chonepm$ mediated LSP pair annihilations 
to  $W^+W^-$ to play a significant role. Annihilation into ZZ, Zh
and $t \bar t$ through virtual Z exchange open up for higher LSP masses and  
also contribute significantly. In contrast, the mechanism for satisfying relic density limits 
in the similar branch
of the {\it wino model} is mainly due to  coannihilations of the LSP with 
stau/sneutrino
(see Fig.~4a of Ref. \cite{older}). It is interesting to note that the DM 
constraint also provides with bounds on the sparticle masses from above since
the upper branch ends abruptly above a certain value of $\mlspone$. 
This is due to the fact that for large $\mchonepm$, the chargino mediated 
LSP pair annihilation cross-section is suppressed. In addition, the annihilation
cross-section to $t \bar t$ pairs goes as ${(\frac {m_t}{\mlspone})}^2$\cite{dm_rev1,dmrd_ffbar},
which becomes inefficient due to increasing value of 
$\mlspone$, resulting into over-abundant DM.

The DM relic density satisfied lower branch forming almost a line parallel to the 
$\mchonepm$ axis corresponds to LSP 
pair-annihilation via s-channel Z/h resonance and it mostly falls 
on the 3$\sigma$ zone of the $\gmin2$ constraint as discussed above. Moreover,
a large portion of this line is excluded by slepton searches for relatively 
low $\mchonepm$. The constraints from slepton searches are considerably relaxed
in the LHLRS-$\chonepm$ model as discussed in Sec.~\ref{sec:LHLRS}.
 
Thus, novel LHC signatures via the decay $\lsptwo \ra h \lspone$ is a 
potential discovery channel at the LHC (see Sec. \ref{Sec:ConstraintsGluino})
provided we accept the 3$\sigma$ level of agreement with the $\gmin2$ data.

In Fig.~\ref{MH30LR} we present the results for $\tan\beta = 30$. A 
larger value of 
$\tan\beta$ causes the lighter stau (${\tilde \tau}_1$) to become
 the next to the lightest supersymmetric particle (NLSP) due to larger L-R mixing. 
We note that in comparison to a similar wino-dominated 
$\chonepm$ scenario like what appears in Fig.~4b of Ref.\cite{older},
the theoretically discarded region corresponding to 
low $\mlspone$ is smaller in Fig.~\ref{MH30LR}.
It is partly due to the
reasons already discussed for Fig.~\ref{MH6LR}.  
Moreover, $\mu$ being relatively low, as demanded by a higgsino 
dominated $\chonepm$,  the mixing in the stau sector is suppressed.
As a result, the portion of the parameter space with stau as the LSP 
is smaller.  No $\mchonepm = \mlspone$ line exists 
in this case as ${\tilde \tau}_1$ becomes the LSP for 
$\mchonepm \sim \mlspone$ because of large mixing.
Fig.~\ref{MH30LR} shows further that 
the LHC exclusion contour from the trilepton data is too small to be of any importance in
determining the APS. In fact, the LHC limits are fully encompassed 
by the theoretically excluded region. In addition to the reasons already discussed, 
the constraints are also weakened because of 
a significantly reduced 
mass of ${\tilde \tau}_1$ in comparison to the same of 
sleptons of the first two 
generations. Consequently, there is a significant suppression in the BRs 
of the decays of $\chonepm$, $\lsptwo$ and $\lspthree$ into states involving $e$ and/or $\mu$.
This explains the shrinkage of the LHC forbidden region 
compared to Fig.~\ref{MH6LR}.

The $\gmin2$ constraint that typically 
scales with $\tan\beta$ is satisfied much easily in this large 
$\tan\beta$ scenario even at 1$\sigma$ level.
The largest contribution to $\gmin2$ is from
the neutralino-smuon loop. But the chargino-sneutrino loop also 
has a significant contribution.

The features of the upper branch allowed by the WMAP/PLANCK data
are similar to those in Fig.~\ref{MH6LR}.
However, since ${\tilde \tau}_1$ is the NLSP in this case
stau coannihilation also becomes important
for higher LSP masses. Nevertheless,  the upper branch is 
also truncated in this case, thereby imposing an upper 
limit on the sparticle spectra. This bounded region is 
consistent with both the $\gmin2$ and the LHC constraints. 
We recall that in the wino dominated scenario 
(see Fig.~4b of Ref.\cite{older}),
the Higgs resonance region disappears 
at large $\tan\beta$ due to vanishingly small higgsino 
component of $\lspone$ which severely suppresses the
h$\lspone\lspone$ coupling. 
In contrast, $\mu$ being smaller in the present scenario, 
$\lspone$, though dominantly a bino, has sufficient higgsino components.  
Thus, the h-resonance region extends to large values of 
$\mchonepm$. The bulk of this region, though allowed by 
the $\gmin2$ constraint, is disfavoured by the direct slepton search data. 
However, as already noted,  this region opens up in the LHLRS$_{\chonepm}$ model.\\

\begin{figure}[!htb]
\vspace*{-0.05in}
\mygraph{MM6LR}{MM6LR.eps}
\hspace*{0.5in}
\mygraph{MM30LR}{MM30LR.eps}
\caption{Plots in the $\mchonepm-\mlspone$ plane for the LMLRS scenario 
with $\tan\beta$=6 and 30. Colours and conventions are the same as in Fig.~\ref{MH_LR}.}
\label{MM_LR}
\end{figure}

\subsection {The Light Mixed $\chonepm$ with light Left and Right Sleptons (LMLRS) model}
\label{sec:LMLRS}
The analysis of  this subsection is based on the {\bf L}ight 
{\bf M}ixed $\chonepm$ with light {\bf L}eft and {\bf R}ight 
{\bf S}leptons (LMLRS) scenario
where the states $\chonepm$, $\lsptwo$ and $\lspthree$ are  admixtures of large wino and higgsino 
components while the slepton masses are chosen as in the previous subsection. 
We consider $\mu = 1.05~M_2$, a choice that corresponds 
to a large higgsino-wino mixing while keeping $\chonepm$, $\lsptwo$ and 
$\lspthree$ to be still dominated by the higgsinos. 
On the other hand, 
$\lspone$ may have a significant amount of higgsino component 
in some region of parameter 
space, although it is by and large a bino-dominated state ($\mu > M_1$).
We analyze the production processes : $p p \ra \chonepm~\lsptwo/\lspthree$. 
In Fig.~\ref{MM6LR}, we show our results for tan$\beta =$~6. 

A gradual weakening of the LHC constraint from the trilepton 
search is evident from the LWLRS model (Fig.~4a of Ref.\cite{older}), to the LMLRS model 
(Fig.~\ref{MM6LR}) and then to the LHLRS model (Fig.~\ref{MH6LR}) 
while the higgsino fraction within $\chonepm$ increases steadily.
This is as expected from the discussions on the 
cross-sections in Sec. \ref{Sec:ProductionElectroweakinos}.

Thus, for a negligible LSP mass the lighter chargino mass limit 
that is allowed via the trilepton search 
is about 550~GeV in Fig.~\ref{MM6LR} compared to about 390 GeV in 
Fig.~\ref{MH6LR}. However, the model independent results of slepton searches 
push the above limits in the figures 
to almost an identical value ($\simeq$600~GeV).  


\begin{table}[!htb]
{\small
\begin{center}
\begin{tabular}{|c|c|c|c|c|}
\hline
Point from Fig. & 2a & 2a & 2b & 4b \\
\hline
$M_1$ & 270 & 59 & 239 &  316 \\
\hline
$M_2$ & 342 & 601 & 348 & 374 \\
\hline
$\mu$ & 359 & 631 & 365 & 393 \\
\hline
$\mlspone$ &  251 & 57 & 228 & 298 \\
\hline
$\mchonepm$ & 301 & 568 & 315 & 342 \\
\hline
$\mlsptwo$ & 311 & 569 & 318 & 351 \\
\hline
$\mlspthree$ &369   &639     &376   &403\\
\hline
$\M_{{\tilde e, \tilde \mu}_L}^D$ & 284 & 443 & 277 & 327 \\
\hline
$\M_{{\tilde e, \tilde \mu}_R}^D$ & 284 & 443 & 277 & 2000  \\
\hline
$m_{\tilde \tau_1}$ & 278 & 435 & 238 & 327 \\
\hline
$M_{\nu}^D$  & 274 & 436 & 266 & 318  \\
\hline
$\Omega_{\tilde \chi} h^2$ & 0.11 & 0.098 & 0.14 & 0.1   \\
\hline
$\sigma_{SI} (pb) \times 10^{-9}$ & 12 & 0.078 & 2.32 & 9.36 \\
\hline
$a_\mu^{\rm SUSY} \times 10^{-10}$ &  6.5 & 2.2 & 34  & 25 \\
\hline
\end{tabular}
  \end{center}
    \caption{The sparticle spectra corresponding to different points
chosen from  Fig.~\ref{MM_LR} and Fig.~\ref{MM_L}. All the masses are in GeV.}
\label{tab2}
}
\end{table}

\begin{table}[!htb]
{\small
\begin{center}
\hspace{-1 cm}
\begin{tabular}{|c|c|c|c|c|c|c|c|c|c|}
\hline
\hline
Decay Modes&\multicolumn{4}{c|}{Point from Fig.}&Decay Modes&\multicolumn{4}{c|}{Point from Fig.}\\
\cline{2-5}
\cline{7-10}
			& 2a & 2a & 2b & 4b &        & 2a & 2a & 2b & 4b \\
\hline
\hline
$\lsptwo  \ra  \slepl^\pm l^\mp       $&38.4       &30    &26     &44    &  &  &  &  &  \\
$\quad  \ra\snu \bar{\nu}      $  &24    &42  &28.8   &30    &  & & & &  \\
$\quad  \ra  \slepr^\pm l^\mp       $&10      &-   &2.4    &-    &  & & & & \\
$\quad  \ra  \stauonepm \tau^\mp    $&18      &10    &42    &26   & $\lspthree  \ra  \slepl^\pm l^\mp       $&0.6       &-     &-       &0.4 \\
$\quad \ra {{\wt\tau_2}^\pm}\tau^\mp$&7.4     &6   &        &-  & $\quad  \ra\snu \bar{\nu}      $&5.4       &1  &3.6    &6     \\ 
$\quad  \ra  \lspone h       $&-       &9     & -      &-  & $\quad  \ra  \slepr^\pm l^\mp       $&0.8       & -    &-      &- \\
$\quad  \ra  \lspone Z      $&-       &3   &       &- &$\quad  \ra  \stauonepm \tau^\mp    $&0.6    &0.8  &34     & 50\\
\cline{1-5}
\cline{1-5}
$\chonepm    \ra \snutau \tau       $&25    &16    &20   &31  & $\quad \ra {{\wt\tau_2}^\pm}\tau^\mp$&3.4       &2  &15      &- \\
$\quad    \ra\stau_{1} \nutau  $&9     &8.5     &32   &7 & $\quad  \ra  \lspone h$      &          &24    &0.7     &-\\
$\quad    \ra \stau_{2} \nutau $&1     &5   &-  &-    &  $\quad  \ra  \lspone Z$    &89        &71   &45  &42   \\
$\quad    \ra\snu_l l   $ &50    &32    &32   &48   &  &  &  &  & \\
$\quad    \ra \wt{l}_L \nu_{l}$ & 16    &26  &14.8 &14  &  &  &  &  & \\
$\quad    \ra W^{\pm} \lspone$  &-     &12   &1.4  &  &  &  &  &  &  \\

\hline
\end{tabular}
 \end{center}
  \caption{ {The decay modes and BRs of different points taken from Fig.~\ref{MM_LR} and Fig.~\ref{MM_L}. } }
\label{tab3}
}
\end{table}
 

We observe that in the small $\mchonepm$ region of Fig.~\ref{MM6LR},  
the limits from the LHC trilepton searches are
practically unaltered with respect to the corresponding {\it wino model} 
(Fig.~4a of Ref.\cite{older}), in spite of differing wino and higgsino components
within $\chonepm$ between the two scenarios. 
Here BR of $\lsptwo$ decaying into invisible   
$\nu {\tilde \nu}$ final state reduces significantly due to an increase in its higgsino fraction. 
This on the other hand is compensated the reduction in the cross-section via an increase in the leptonic BR 
leading to an almost unchanged LHC limit.
We further note that
$\lspthree$ does not contribute to the trilepton signal appreciably 
since
it decays principally via $Z \lspone$ mode 
with BR around 80\%. Moreover, the production cross-section 
for $\lspthree~\chonepm$ is rather small
compared to $\lsptwo~\chonepm$ since 
$\mlspthree$ is somewhat larger than $\mlsptwo$. 
Some of the above features are illustrated by two  
representative points taken from Fig.~\ref{MM6LR}) in Tables \ref{tab2} 
and \ref{tab3}. In the region of high $\mchonepm$, the decay modes 
$\lsptwo \ra h \lspone $ and $\chonepm \ra W^{\pm} \lspone$ open up 
(the BRs lie around 10\%). $\lspthree$ decays dominantly into
 $Z \lspone$ as before but a sizable fraction goes to $ h \lspone$.  
Thus, the LHC exclusion contour shrinks when compared  with Fig.~4a of Ref.\cite{older}. 
These features are illustrated in Table \ref{tab3}.

Regarding $\gmin2$, in all the cases the LHC allowed parameter space is
consistent at 3$\sigma$ level.
Both the loops involving neutralinos as well as charginos contribute with comparable magnitudes.

Along the upper red dotted line,
the main contribution to the observed relic density are chargino mediated 
LSP annihilation into the final state $ W^{+}W^{-}$ and
LSP pair annihilation into  $t \bar t$ . Some amount of bulk annihilation is also present for
low $\mchonepm$ although this region of parameter space is discarded by the LHC constraints.  
Unlike the {\it higgsino model} of
Fig.~\ref{MH6LR} the upper branch does not end abruptly (at  $\mchonepm \approx 400$~GeV).  This is due to the fact
that $\lspone-\chonepm-W^{\pm}$ coupling which is behind the $\chonepm$ mediated annihilation to  $ W^{+}W^{-}$
is stronger  because of larger wino content of the LSP in the {\it mixed model}. 
In the corresponding {\it wino model} (Fig.~4a of Ref. \cite{older}), however,
the relic density falls within the observed range 
mainly due to stau/sneutrino coannihilations.
The lower branch of red points consistent with the observed DM relic 
density arises through Z/h resonance.

Fig.~\ref{MM30LR} illustrates the results for tan$\beta =$ 30.  
The small change in the theoretically discarded region  is due to reasons explained
earlier (see Sec. \ref{sec:LWLRS}). Here the parameter space excluded by the LHC constraint
shrinks with respect to that in the {\it wino model} (Fig.~4b of Ref. \cite{older}). 
The reasons are the same as the ones discussed in the context of 
Fig.~\ref{MM6LR} and \ref{MH30LR}. However, $\lspthree$ decay has no bearing  on 
the LHC  limits in this case. Significantly larger parameter spaces consistent 
with the $\gmin2$ constraint at the level of 
1$\sigma$ and 2$\sigma$ are available in this high $\tan\beta$ scenario. 
On the other hand, for the DM constraint as satisfied in the the upper red dotted
branch, the main DM producing mechanisms are stau-LSP coannihilation and  
stau-stau annihilation as in the {\it wino model} (Fig.~4b of Ref. \cite{older}).
Chargino-mediated annihilation and annihilation into 
$t\bar t$ pairs are also present albeit to a lesser extent.
The features of one representative point in this parameter space
is illustrated in Tables \ref{tab2} and \ref{tab3}.
Here the Higgs resonance strip extends to higher $\mchonepm$ compared to the 
{\it wino model} (see Fig.~4b of Ref.\cite{older})
where it was practically absent.  This is due to a modest increase in the higgsino component of the LSP. 
However, this parameter space is strongly disfavoured by the LHC limits even in the 
LMLRS$\chonepm$ model. 

We end this section by noting that in both the LHLRS and LMLRS models with high 
$\tan\beta$ we obtain several APSs consistent with the main constraints.  


\section{The Higgsino and Mixed Models with Light Left Sleptons}
\label{Sec:HiggsinoAndMixedLS}
\subsection{A brief review of the Light Wino and light Left Sleptons (LWLS) model}
We recall that in Fig.~1 of our previous analysis (Ref. \cite{older}),
the model characterized by {\bf L}ight {\bf W}ino and light {\bf L}eft 
{\bf S}leptons (LWLS) yields the strongest mass limit on $\chonepm$. 
For negligible $\mlspone$, $\mchonepm$ ranging up to 
610 GeV is excluded. In all cases the low $\tan\beta$ scenarios
are consistent with the $\gmin2$ constraint at best at the 3$\sigma$ 
level. On the other hand, for large $\tan\beta$, the SUSY prediction has a 
better agreement with constraint which is satisfied at 1$\sigma$ or 2$\sigma$ levels. 
In all cases the parameter space consistent with the WMAP/PLANCK 
data has two limbs.  In the upper limb various coannihilations 
lead to the desired relic density. The lower limb which represents the LSP
pair annihilation through the h-resonance is either absent or 
strongly disfavoured by the LHC constraints in the high $\tan\beta$ scenarios.

\subsection {The Light Higgsino and light Left Sleptons (LHLS) model}
\label{sec:LHLS}
In this section  we replace {\it wino} of the previous subsection by a 
{\it higgsino} and
focus on models with {\bf L}ight {\bf H}iggsino and light {\bf L}eft 
{\bf S}leptons (LHLS)
where $\chonepm$ and $\lsptwo$ are higgsino-dominated and left 
slepton masses are kept midway between $\mchonepm$ and $\mlspone$.
Right sleptons are taken to be heavy.

Fig.~\ref{MH6L} shows our results for the model
with $\tan\beta$=~6.
If we compare this with Fig.~1a of Ref.~\cite{older} we see that
the theoretically discarded regions are almost the same in both cases.  The small change 
is due to a different choice of EWSB scale as noted earlier.  The LHC  limits
from the trilepton searches are significantly degraded in the {\it higgsino} case.
The reasons are the same as the ones in Sec. \ref{sec:LHLRS}.
For low LSP masses, however, one finds a region with $\mchonepm \gsim$ 600~GeV 
to be allowed from slepton search. This limit will be relaxed to 410~GeV in the tilted 
LHLS$_{\chonepm}$ models (See \cite{older} Figs. 3a and 3b).

The major contribution to $\gmin2$ comes from the 
chargino-sneutrino loop as  expected in models with 
light L-type sleptons (see Ref. \cite{older} for details).
The $\gmin2$ allowed regions at the level of 1$\sigma$, 2$\sigma$ and 3$\sigma$ are similar to
those of Fig.~1a of Ref.~\cite{older}.  The observations regarding the points satisfying the WMAP/PLANCK limits
are similar to those of Fig.~\ref{MH6L} i.e. the lower red dotted branch comes from Z/h resonance 
annihilation and the upper branch from annihilation to  $ W^{+}W^{-}$, $t \bar t$, $ZZ$ 
pairs. There is a small amount of sneutrino coannihilation in the upper branch along with a little bulk annihilation
for very low $\mchonepm$ disfavoured by the LHC bounds.  
The reasons for the abrupt end of the upper branch indicating upper bounds on $\mchonepm$ and $\mlspone$ are
already discussed in Sec. \ref{sec:LHLRS}. 
The $\gmin2$ constraint is satisfied only at the level of 3$\sigma$.\\

\begin{figure}[!htb]
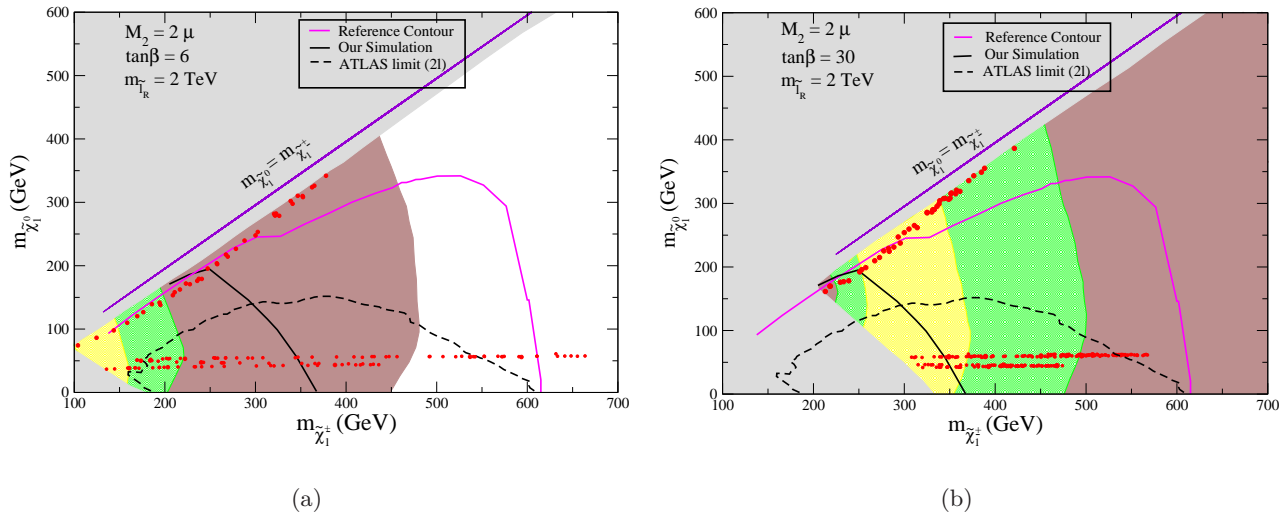

\vspace*{-0.05in}
\mygraph{MH6L}{MH6L.eps}
\hspace*{0.5in}
\mygraph{MH30L}{MH30L.eps}
\caption{Plots for the LHLS model
with $\tan\beta$=~6 (a) and 30 (b).  Colours and conventions are the 
same as in Fig.~\ref{MH6LR}.
}
\label{MH_L}
\end{figure}

Our results for $\tan\beta~$=~30 is shown in Fig.~\ref{MH30L}.  
The collider exclusion limit is the same 
as that of Fig.~\ref{MH6L}.  Since the right sleptons are heavy, 
there is no large mixing in the stau sector. As a result,
the LHC exclusion contour remains unaltered in spite of the 
change in $\tan\beta$. The dominant contribution
to $\gmin2$ comes from the chargino-sneutrino loop diagram
and it is characteristically similar to Fig.~\ref{MH6L} except that
there is the usual $\tan\beta$ enhancement leading to valid 1$\sigma$ and 2$\sigma$ regions.
Annihilation and coannihilation mechanisms of the LSP in the regions satisfying 
the WMAP/PLANCK constraints in this scenario are similar to those of Fig.~\ref{MH6L}. The 
LSP pair annihilation into the Higgs resonance is allowed in the LHLS$\chonepm$ model.

\subsection {The Light Mixed $\chonepm$ and light Left Sleptons (LMLS) model}
\label{sec:LMLS}
In Fig.~\ref{MM6L} we show our results for the {\bf L}ight 
{\bf M}ixed $\chonepm$ and light {\bf L}eft {\bf S}leptons model
(LMLS) with $\tan\beta$=~6.
The collider exclusion contour is weakened compared to the reference contour of 
Fig.~1a of Ref.\cite{older}.  The reasons are almost same as those mentioned in the 
discussions for  Fig.~\ref{MM6LR}.
Major contribution to $\gmin2$  is provided by the chargino-sneutrino loop diagram.  
The points satisfying the WMAP/PLANCK limits
in the lower branch arise as a result of
 Z/h resonance annihilation processes. For the upper
branch, main mechanisms are  annihilations into $W^{+}W^{-}$, $ZZ$, $t \bar{t}$ pairs.\\

\begin{figure}[!htb]
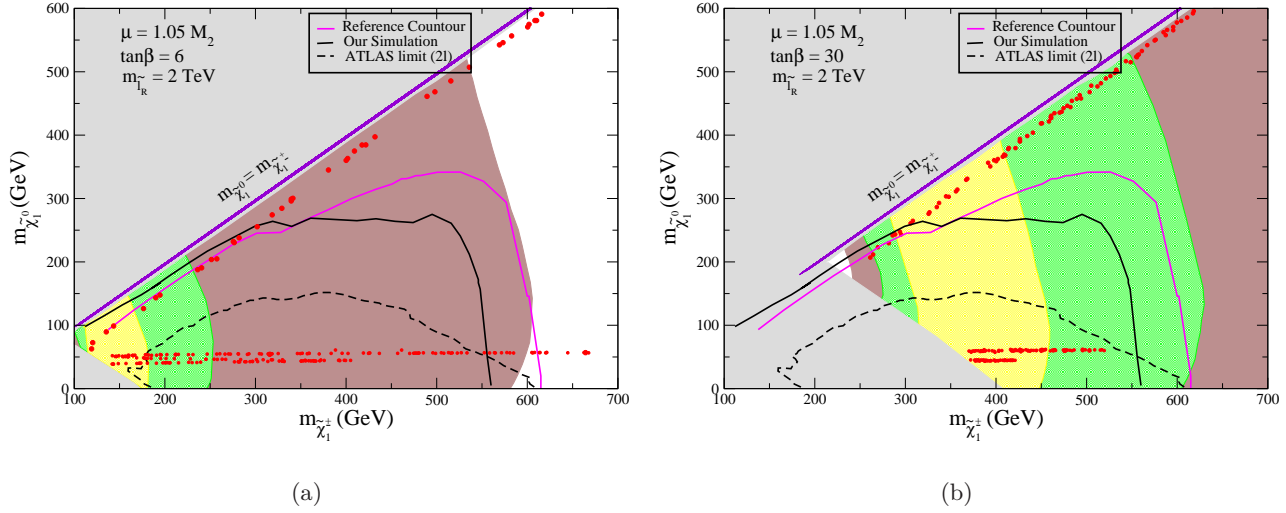

\vspace*{-0.05in}
\mygraph{MM6L}{MM6L.eps}
\hspace*{0.5in}
\mygraph{MM30L}{MM30L.eps}
\caption{Plots for the LMLS model
with $\tan\beta$=6 (a) and 30~(b).  Colours and conventions are the same as in Fig.~\ref{MH6LR}.
}
\label{MM_L}
\end{figure}

Fig.~\ref{MM30L} shows the results for large $\tan\beta$ in the LMLS scenario.
The observations regarding the upper limb
consistent with the WMAP/PLANCK constraints are 
similar to those of Fig.~\ref{MM6L}. 
As in all high $\tan\beta$ scenarios there are portions of the parameter
space consistent with the $\gmin2$ constraint at 1$\sigma$ and 2$\sigma$ levels 
leading to an APS consistent with all the major constraints.
We show the features of one representative point for this scenario 
in Tables \ref{tab2} and \ref{tab3}.\\


\section{The Light Higgsino and Heavy Sleptons (LHHS) model}
\label{sec:LHHS}

\begin{figure}[!htb]
\begin{center}
\includegraphics[scale=0.35]{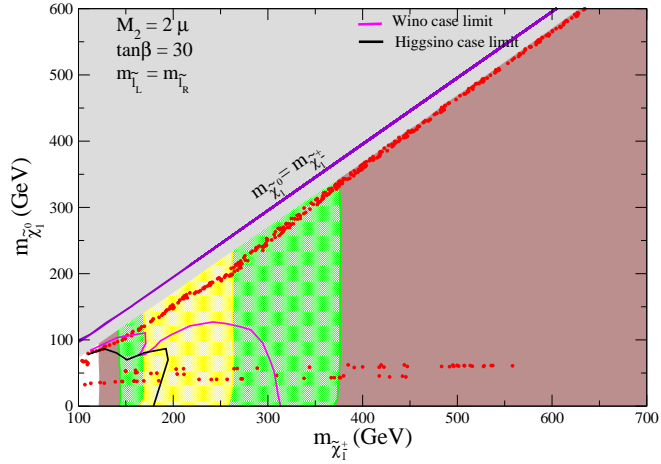}
\end{center}
\caption{Plot for the LHHS scenario with tan$\beta=$30. The magenta line represents the exclusion
contour at 8 TeV by the ATLAS collaboration\cite{atlas3lew} for the {\it wino model}. The black
line stands for the much weaker exclusion contour in the corresponding {\it higgsino model}
obtained by our simulations. Other colours and conventions are the same as in Fig.~\ref{MH6LR}.}
\label{LHHS}
\end{figure}

In the {\bf L}ight {\bf H}iggsino and {\bf H}eavy {\bf S}leptons (LHHS) scenario 
the chargino decays to W$\lspone$ with 100\% BR and the
decay  $\lsptwo \ra Z \lspone$ has 100\% BR for $(\mlsptwo-\mlspone) < m_h$.
For larger $\mlsptwo$, the mode $\lsptwo \ra h \lspone$ opens up and  
its BR can be $\approx$ 40\% near threshold.  With  increasing value of $\mlsptwo - \mlspone$,
the BR($\lsptwo \ra h \lspone)$ may be as large as 60\%. The presence of this 
mode is the main difference with the LWHS scenario and leads to the ($ W + h + \met$) signal. 
The BR of the mode $\lspthree \ra Z \lspone$ is around 70\%.
However, the LHC limits in this  case are degraded due to a reduction in cross-section as
already discussed. For $\mchonepm >$ 200 GeV, all the LSP masses are allowed.
On the other hand, for negligible LSP masses, a weaker limit ($\mchonepm \gsim$ 175 GeV) is obtained.

The above features, and, consequently the LHC exclusion contours, 
are fairly independent of $\tan\beta$. However, as already shown in the 
last two sections, the constraints from $\gmin2$ is effective for high 
$\tan\beta$. This is particularly true in a situation with 
heavy slepton masses that potentially reduces $\amususy$.  
Thus, only the case of $\tan\beta$ = 30 
is shown in Fig.~\ref{LHHS}. The SUSY contribution to $\gmin2$ is
dominated by the chargino-sneutrino loop diagram. The upper branch of 
the region consistent with the WMAP/PLANCK data arises
due to chargino-mediated LSP pair annihilation to W pairs. There is also annihilation 
to fermion anti-fermion pairs
through virtual Z exchange and some amount of LSP-chargino coannihilation. 
The Z/h-resonance region is also allowed in this scenario by the major constraints and the
$\gmin2$ constraint is satisfied even at the level of 1$\sigma$. 

An analogous case of a Light Mixed and Heavy Sleptons (LMHS) scenario would have 
the corresponding LHC
exclusion contour lying in between the magenta and the black lines of Fig.~\ref{LHHS}.
This is expected from the chargino-neutralino production 
cross-sections as discussed in Sec.~\ref{Sec:ProductionElectroweakinos}. 
We do not present the details here since no qualitatively new feature of 
the signal emerges from this analysis.

\section{Direct Detection via Spin-independent scattering}
\label{Sec:DirectDMSIDetection}
Spin-independent (SI) interaction of the lightest neutralino with 
quarks inside the detector nucleus 
occurs via s-channel squark exchange and t-channel Higgs exchange processes. 
With the present LHC bounds squarks are considerably heavy. Hence, the 
Higgs exchange diagrams would dominantly contribute towards
$\sigma_{p\chi}^{SI}$, the spin-independent ${\tilde \chi} -p$ scattering 
cross-section\cite{Drees:1993bu}. 
In this context 
we note that the $h(H) \lspone \lspone$ coupling involves products of the 
gaugino and the higgsino components of the neutralino diagonalizing 
matrix\cite{SUSYbooks}.  
Unless $|{M_1-|\mu|}| \lsim M_Z$, in 
the Higgs decoupling zone\cite{SUSYbooks} satisfying $m_H \simeq m_A >>M_Z$, 
the Higgs couplings to a bino-dominated LSP is approximately 
given as below\cite{Hisano:2009xv}.  

\begin{eqnarray}
C_{h \tilde \chi \tilde \chi} &\simeq& \frac{m_Z s_W t_W}{M_1^2 - \mu^2}
    \bigl[ M_1 + \mu \sin2 \beta \bigr],\nonumber\\
C_{H \tilde \chi \tilde \chi} &\simeq& - \frac{m_Z s_W t_W}{M_1^2 - \mu^2}
    \mu \cos2\beta .
    \label{SIhiggscoup}
\end{eqnarray}

\noindent
Here $s_W = \sin \theta_W$ etc. with $\theta_W$ as the Weinberg angle.
Similar results for a wino or a higgsino-dominated LSP may be seen in 
Ref.~\cite{Hisano:2009xv}.
Clearly, the above shows that the 
SI cross-section $\sigma_{p\chi}^{SI}$ 
would be large when there is a significant amount of 
bino-higgsino mixing i.e. $M_1 \simeq \mu$. This is unlike   
a pure gaugino or a pure higgsino DM when 
the associated SI cross-section would become quite small.  
We note that in contrast to our previous analysis\cite{older}, 
where we considered a bino dominated LSP with 
$\mu >> M_1,M_2$, the present work has a significant amount 
of higgsino component in the LSP. As a result, $\sigma_{p\chi}^{SI}$
is typically larger than what was seen in 
Ref.~\cite{older} and in most of the cases its values lie above
the LUX\cite{LUX} limit.
However, we must note that 
there still exists a significant amount of uncertainty in the theoretical
estimate of the SI cross-section (for a brief discussion see 
Ref.~\cite{older} and references quoted therein). This at least
relates to issues like  uncertainties in the determination of the strangeness
content of nucleon, local DM density and velocity distribution profiles. 
All these uncertainties may accommodate lowering of the cross-section 
by an order of magnitude.
We will present our 
results in the following subsections for the {\it higgsino} and the {\it mixed models}.
We like to point out that no tilted scenarios have been 
included in our analyses on direct and indirect detection of dark matter.

\subsection{LHLRS and LMLRS}
\noindent
The results for the SI direct detection are shown in Fig.~\ref{dd_MHLR} 
and Fig.~\ref{dd_MMLR} corresponding to i) higgsino-dominated 
$\chonepm$ (LHLRS) and ii) wino-higgsino mixed $\chonepm$ (LMLRS)
analyses of Fig.~\ref{MH_LR} and Fig.~\ref{MM_LR} respectively. 
Fig.~\ref{dd_MHLR} combines the results in the $\mlspone-\sigma_{p\chi}^{SI}$
plane corresponding to two values of $\tan\beta$ (blue and cyan  
for $\tan\beta=6$ and 30 respectively) as used in Fig.~\ref{MH6LR} and 
Fig.~\ref{MH30LR}. Similarly, Fig.~\ref{dd_MMLR} details the results 
corresponding to Fig.~\ref{MM6LR} and Fig.~\ref{MM30LR} for the 
same values of $\tan\beta$ as mentioned above. 
Figs.~\ref{dd_MHLR} shows only the allowed points that satisfy the relic density 
limits, the $\gmin2$ constraint 
(upto 2$\sigma$ for large $\tan\beta$, and 3$\sigma$ for small 
$\tan\beta$ cases) and the collider limits 
for $\mlspone$ between 200 and 350 GeV. $\sigma_{p\chi}^{SI}$
for the points 
exceed the LUX limit, while staying
within an order of magnitude of the same limit\footnote{This is especially so for models
with high $\tan\beta$ which are also in better agreement with the $\gmin2$ constraint.}. 
In regard to the {\it mixed model} (Fig.~\ref{dd_MMLR}), 
we similarly find that parameter zones satisfying $\mlspone > 200$~GeV are consistent with the 
major constraints.  Here also the points
exceed the LUX limit by a similar 
amount. This is unlike the analysis of 
the LWLRS scenario of Ref.~\cite{older} where it was not difficult to 
satisfy the LUX limit.  Since the parameter points correspond to a 
deviation below an order of magnitude from the LUX limit, we consider 
them to be presently acceptable in view of the uncertainties discussed before.
There are a few points in the Higgs resonance region for the low $\tan\beta$ case.
Since Higgs-pole annihilation occurs for $\mlspone \sim \frac{M_h}{2}$ 
and the allowed points from this region correspond to very high $\mchonepm$ 
(see Fig.~\ref{MM6LR}), the LSP in this region is highly bino-dominated,
thus having very small values of $\sigma_{p\chi}^{SI}$.
The XENON1T experiment will conclusively probe these models.\\

\begin{figure}[!htb]
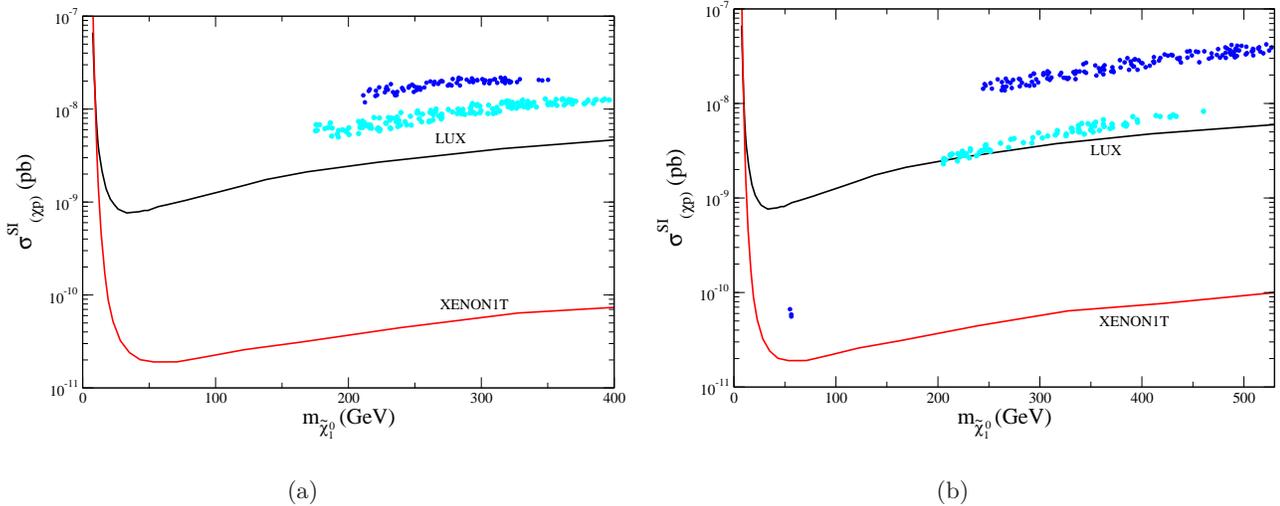

\vspace*{-0.05in}
\mygraph{dd_MHLR}{dd_MHLR.eps}
\hspace*{0.5in}
\mygraph{dd_MMLR}{dd_MMLR.eps}
\caption{Direct detection results for the LHLRS (a) and LMLRS (b) scenarios
with $\tan\beta$ = 6 (blue) and 30 (cyan).
The LUX and XENON1T limits are shown as black and red lines respectively.  The points
satisfying the WMAP/PLANCK and LHC limits
are shown. The $\gmin2$ constraint is applied up to the level of 
3$\sigma$ for low and 2$\sigma$ for high $\tan\beta$ cases respectively. 
}
\label{dd_LR}
\end{figure}

\subsection {LHLS and LMLS}

\begin{figure}[!htb]
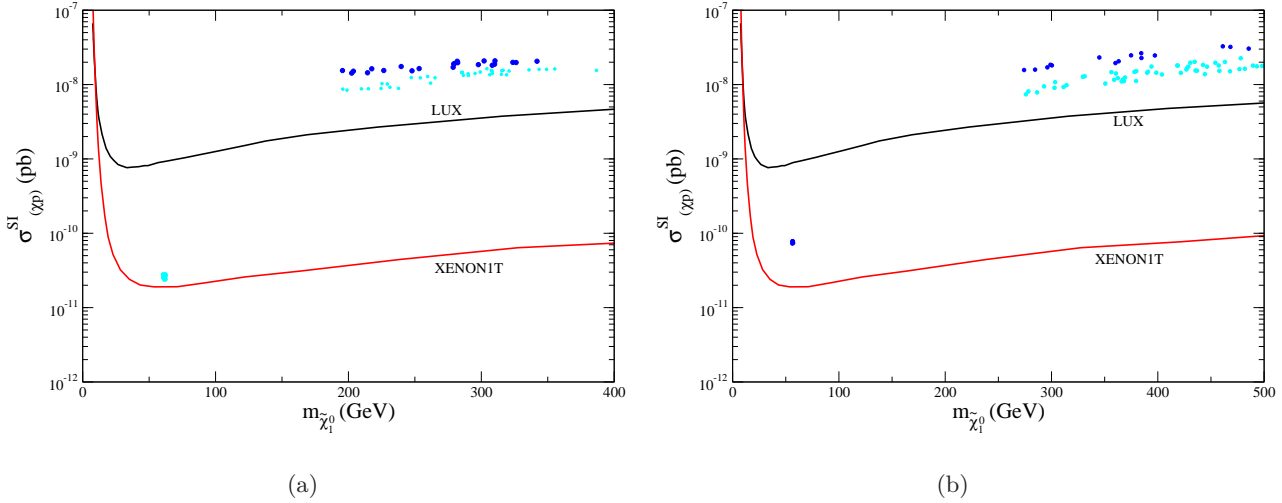

\vspace*{-0.05in}
\mygraph{dd_MHL}{dd_MHL.eps}
\hspace*{0.5in}
\mygraph{dd_MML}{dd_MML.eps}
\caption{Direct detection results for the LHLS (a) and LMLS (b) scenarios with 
$\tan\beta$ = 6 (blue) and 30 (cyan). Colours and conventions are the same as in 
Fig.~\ref{dd_LR}. 
}
\label{dd_L}
\end{figure}

The results for the SI direct-detection are shown in Fig.~\ref{dd_MHL} 
and Fig.~\ref{dd_MML} corresponding to the LHLS and LMLS scenarios for the
analyses of Fig.~\ref{MH_L} and Fig.~\ref{MM_L} respectively. 
The colours (blue and cyan for low and high values of $\tan\beta$ as used in 
Figs.~\ref{MH6L} and \ref{MH30L} respectively) and conventions used in Fig.~\ref{dd_MHL}
are the same as in Fig.~\ref{dd_MHLR}.
Fig.~\ref{dd_MHL} shows only the points satisfying the major constraints,
which in this case, lie in the range $200$~GeV$ < \mlspone < 350$~GeV. 
On the other hand, Fig.~\ref{dd_MML}, which corresponds to Fig.~\ref{MM6L} 
and Fig.~\ref{MM30L}, shows the above type of allowed points for $\mlspone>300$~GeV. 
These points give rise to $\sigma_{p\chi}^{SI}$
 the LUX limit by a similar amount as before.
There are a few points in both the figures, lying in the Higgs resonance 
region which are still allowed
by all the major constraints. This region of parameter space, having very small values of
$\sigma_{p\chi}^{SI}$, will be fully explored by the XENON1T experiment in future.

\subsection{LHHS}
\begin{figure}[!htb]
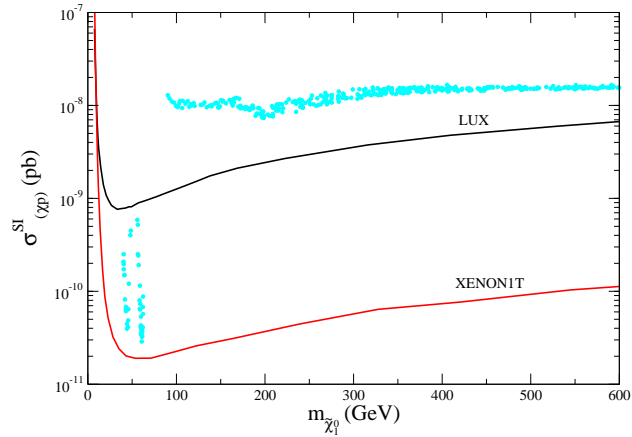

\vspace*{-0.05in}
\begin{center}
\mygraph{dd_HSfig}{dd_hs.eps}
\caption{Direct detection results for the LHHS scenario with 
$\tan\beta$ = 30 for 
higgsino dominated $\chonepm$ scenario.  
The LUX and XENON1T limits are shown as black and red lines. The points allowed by the 
WMAP/PLANCK and collider constraints
are shown. The $\gmin2$ constraint is applied up to the 
level of 3$\sigma$.}
\label{dd_HS}
\end{center}
\end{figure}

The results for the SI direct-detection are shown in Fig.~\ref{dd_HS} 
corresponding to a scenario of a higgsino-dominated 
$\chonepm$ for the LHHS analysis of Fig.~\ref{LHHS} with $\tan\beta=30$. 
In the LHHS analysis we consider masses of the left and the right 
sleptons to be larger than $\mchonepm$ by 200 GeV while a choice of $M_2=2\mu$ 
is made to have higgsino-domination in $\chonepm / \lsptwo$. The LSP 
is however bino-dominated because of the choice $M_1<\mu$. 
Fig.~\ref{dd_HS} shows only the allowed points that satisfy the main constraints.
Parameter points with $\mlspone$ above 80 GeV satisfy these criteria. 
The same is true for a small region around 
Higgs pole annihilation zones i.e. 
$\mlspone \simeq \frac{M_h}{2}$.  
$\sigma_{p\chi}^{SI}$ for the parameter points with $\mlspone$ above 80 GeV exceed 
the LUX limit but they are still within an order of magnitude. 
On the other hand, the allowed points in the Higgs resonance region 
lie much below the LUX limit in a region to be conclusively probed
by the XENON1T experiment in near future.\\

\section{Spin-dependent (SD) direct detection cross-section 
and indirect detection reach for muon flux}
\label{Sec:DirectDMSDandIndirect}
A larger higgsino content of the LSP, as explored in this analysis, 
can potentially be interesting for indirect detection of DM. This is principally due to a 
larger spin-dependent ${\tilde \chi}-p$ scattering cross-section
$\sigma_{p\chi}^{SD}$ which results from a large $Z \tilde \chi \tilde \chi$ coupling 
$C_{Z \tilde \chi \tilde \chi}={|N_{13}^2 - N_{14}^2|}$, where $N_{ij}$ refers to 
the elements of the neutralino diagonalizing matrix. 
The coupling $C_{Z \tilde \chi \tilde \chi}$ is a measure of higgsino asymmetry. 
Thus, on one hand, $\sigma_{p\chi}^{SI}$ is 
already large because $\mu$ and $M_1$ are not too far away from each other. 
On the other hand, $\sigma_{p\chi}^{SD}$ is also 
in the larger side due to increased amounts of higgsinos within the LSP.
The above enhancement in the scattering cross-sections of both the types
in turn causes a loss of energy of a DM particle so that their velocities may go below 
the escape velocity. This results into 
gravitational capture of DM within the dense region of an astrophysical 
object. The captured LSPs then undergo pair annihilations.  
Because of an increased higgsino content within the LSP,  
the pair annihilations may lead to highly energetic 
neutrinos within the Sun. The resulting muon flux out of the charged current 
interactions from the neutrinos coming from the Sun 
may effectively be probed at the IceCube\cite{Icecube1,Icecube2} experiment. 
In SUSY, neutralino annihilations at tree level would not produce 
neutrinos. However, the latter may arise from  gauge bosons, heavy quarks, 
$\tau$-leptons etc. The neutrinos may thus have a broad energy distribution 
with the energy being limited to an appreciable fraction of the mass of the
LSP. Neutrinos from $b \bar b$ or $\tau^+ \tau^-$ are the primary channels 
when the LSP is lighter than $M_W$. However, due to a large threshold,
the above neutrinos may not be suitable for detection.  
For massive neutralinos, LSP pair annihilation would produce gauge bosons, 
top quarks or Higgs bosons. A neutralino having a significant amount of higgsinos may pair 
annihilate to produce gauge bosons. This, in turn, may become a suitable source for 
high energy neutrinos.
    
In order to estimate the capture cross-section in relation to the DM 
annihilation cross-section for the Sun, one considers the time evolution of 
$N$ DM particles,
\begin{equation}
\frac{dN}{dt}= C-C_A N^2.
\label{wimpdepletion}
\end{equation}
Here $C$ measures the rates at which DM particles are captured. $C_A$
relates to the strength of depletion due to DM annihilation.  
The annihilation rate $\Gamma_A$ in turn is related to $C_A$ via 
$\Gamma_A=\frac{1}{2}C_A N^2$\cite{Silk, Gould, Griest:1986}. 

Solution of Eq.\ref{wimpdepletion} results into
$\Gamma_A \equiv \frac{1}{2}C_A N^2=\frac{1}{2}C \tanh^2(t/\tau)$ where
$\tau=1/\sqrt{C C_A}$. Within the MSSM and for 
objects like the Sun (that would correspond to 
large annihilation and large capture rates) one finds that for the 
Solar age of $t=t^\odot=4.5 \times 10^9$~years,
it is justified to assume $t/\tau >>1$. This leads to 
$\Gamma_A=\frac{1}{2}C$, an equilibrium scenario out of 
capture and annihilation\cite{Wikstrom:2009}.  However, this is hardly possible 
for a less massive object like the Earth for which a captured LSP would have a 
much smaller escape velocity and where one has dominance of 
the spin-independent interactions in the DM-nuclear scattering. 
This leads to a weaker indirect detection signal in general\cite{Silk}. 
In contrast to the above, the Sun is a massive object with a much larger 
escape velocity for the LSP. At the same time, both SI and SD cross-sections 
are important for the capture of DM particles within the 
Sun\cite{Chacko, Ibarra:2014}. Suitable models are used to relate
the capture cross-section to SI and SD type of DM-nuclear cross-sections. 
Thus, a measurement of muon flux effectively sets limits on both SI and SD 
cross-sections\cite{Silk, Ibarra:2014}. 
Refs.\cite{Scott:2012mq,Ibarra:2013eba,Catena:2015iea} 
may be seen for further details of setting the above limits and 
the associated degree of model dependence. 

\subsection{LHLRS and LMLRS}
Figs.~\ref{sd_MHLR} and \ref{sd_MMLR} show 
the results of scanning the aforesaid pMSSM 
parameter space in scatter plots of $\sigma_{p\chi}^{SD}$ vs 
$\mlspone$ in the LHLRS and LMLRS models corresponding to 
Fig.~\ref{MH_LR} and Fig.~\ref{MM_LR} respectively. 
Blue and cyan points all of which satisfy the 
relic density, the collider limits and the $\gmin2$ data (up to the level of 
3$\sigma$ for low and 2$\sigma$ for high $\tan\beta$) correspond 
to tan$\beta = $~6 and 30 respectively in each of the figures.
Limits derived from the present as well as future reach of the IceCube
experiment\cite{Icecube1,Icecube2} are shown in 
black and red lines respectively.
The {\it higgsino models} are more sensitive to 
the current IceCube data than the {\it mixed models}, as can be seen from the figures.
Clearly, the final IceCube reach will 
exhaust the parameter space while most of the scatter points in general 
lie at most within an order of magnitude below the presently derived bound from the
same experiment.  Some points for the low $\tan\beta$ scenario in the {\it higgsino model}
are tantalizingly close to the present limit. However, the surviving points 
from the Higgs resonance region for Fig.~\ref{sd_MMLR} lie way below the 
reach of even the final IceCube measurement.\\

\begin{figure}[!htbp]
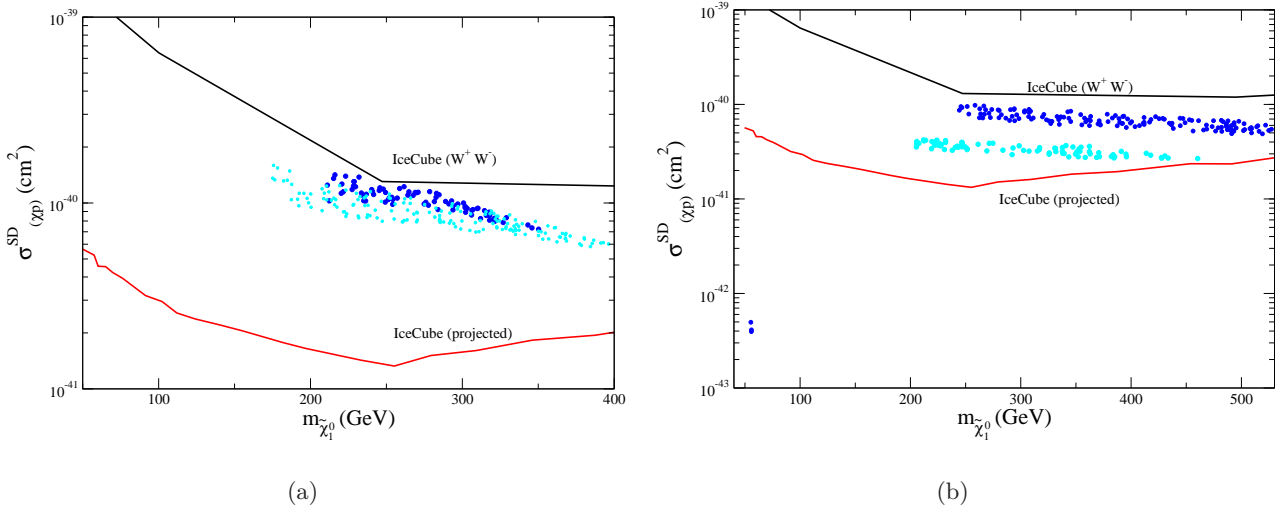

\vspace*{-0.05in}
\mygraph{sd_MHLR}{sd_MHLR.eps}
\hspace*{0.5in}
\mygraph{sd_MMLR}{sd_MMLR.eps}
\caption{Plot of spin-dependent direct detection cross-section 
$\sigma_{p\chi}^{SD}$ vs  the mass of the LSP for 
two cases namely (a) LHLRS :
when $\chonepm$ is principally 
a charged higgsino or (b) LMLRS :
when $\chonepm$ is an appreciable mixture of a charged 
higgsino and wino. Only the points that satisfy the relic density limits, 
the collider 
bounds and the $\gmin2$ data (up to a maximum of 3$\sigma$ level for low and
2$\sigma$ for high $\tan\beta$) are shown.  Blue and cyan points correspond 
to tan$\beta = $~6 and 30 respectively in each of the figures. 
The present and future IceCube limits are shown in black and red lines respectively. 
}
\label{sd_LR}
\end{figure}

Fig.~\ref{muflux_MHLR} and Fig.~\ref{muflux_MMLR} 
show the results of muon flux $\Phi_\mu$ in relation to  
$\mlspone$ in the LHLRS and LMLRS models corresponding to 
Fig.~\ref{MH_LR} and Fig.~\ref{MM_LR} respectively. 
Blue and cyan points, all of which satisfy the major constraints, correspond 
to tan$\beta = $~6 and 30 respectively in each of the figures.
Limits on the muon flux from the present IceCube data and its future 
reach are shown in black and red lines respectively.\\

\begin{figure}[!htbp]
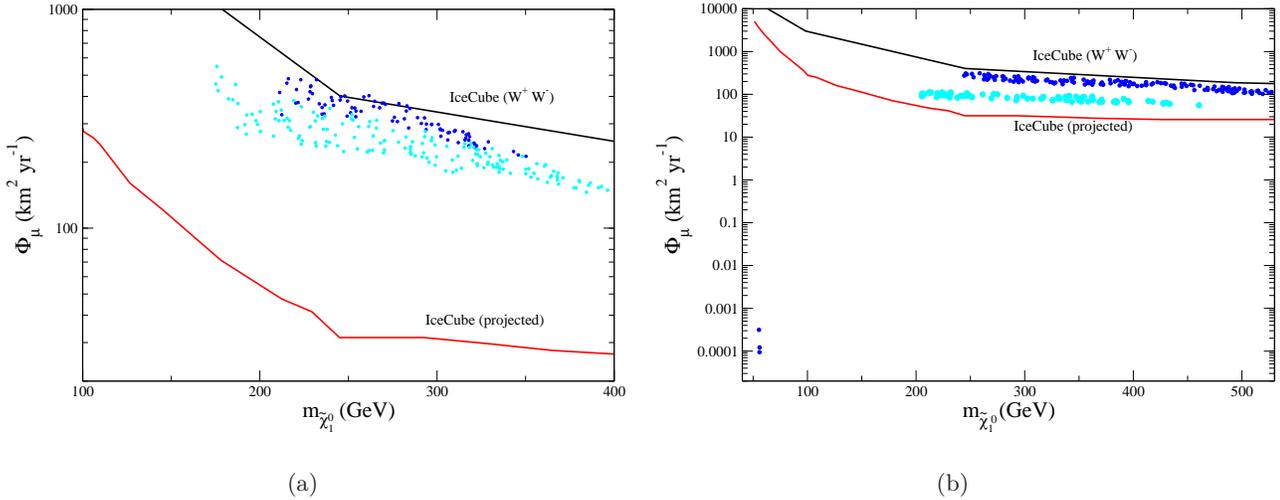

\vspace*{-0.05in}
\mygraph{muflux_MHLR}{muflux_MHLR.eps}
\hspace*{0.5in}
\mygraph{muflux_MMLR}{muflux_MMLR.eps}
\caption{Plot of muon-flux $\Phi_\mu$ vs $\mlspone$ for 
two cases : the (a) LHLRS and (b) LMLRS scenarios.
Only the points that satisfy the relic density limits, the collider bounds 
and the $\gmin2$ data (up to 3$\sigma$ level for low and 2$\sigma$ level for high $\tan\beta$)
are shown.  Blue and cyan points correspond to tan$\beta = $~6 and 30 
respectively in each of the figures. The present and future IceCube limits are shown in 
black and red lines respectively. 
}
\label{muflux_LR}
\end{figure}

The scatter points in general lie within an order of magnitude 
below the presently derived bound from the IceCube. There are some points
in the low $\tan\beta$ case for the {\it higgsino model} which lie very close to the
current data. Clearly, the final IceCube reach will fully explore the
parameter space, except the Higgs resonance region in Fig.~\ref{muflux_MMLR}, 
which is beyond the reach of even the future IceCube measurement. 
There is a high degree of correlation between $\sigma_{p\chi}^{SD}$ and $\Phi_\mu$
for a neutralino with a significant amount of higgsino mixing.
 
\subsection{LHLS and LMLS}
\begin{figure}[!htb]
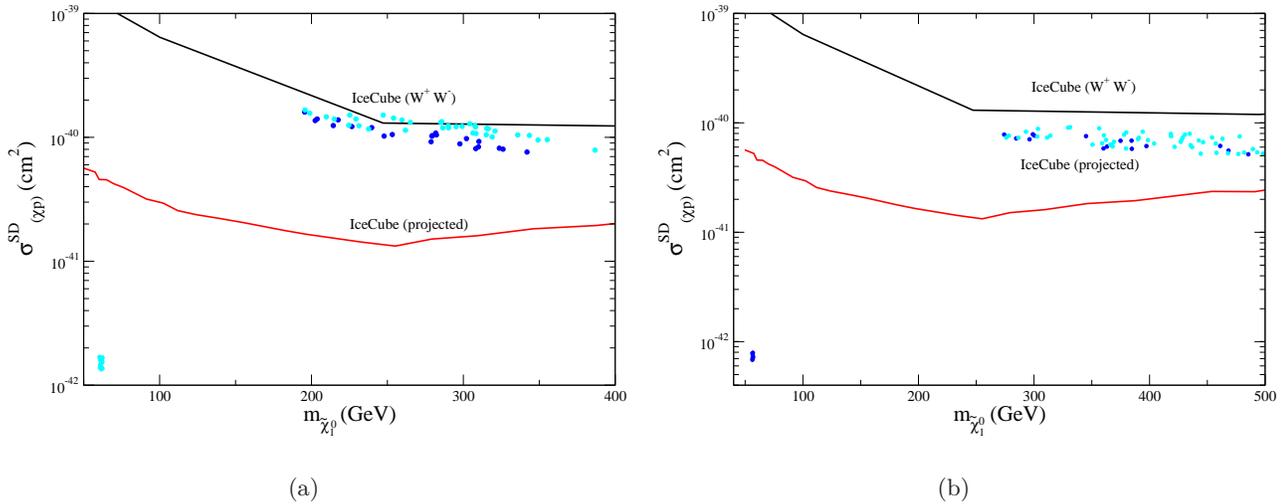

\vspace*{-0.05in}
\mygraph{sd_MHL}{sd_MHL.eps}
\hspace*{0.5in}
\mygraph{sd_MML}{sd_MML.eps}
\caption{Plot of spin-dependent direct detection cross-section vs  
the LSP mass for the two cases : the LHLS (a) and LMLS (b) scenarios.
Colours and conventions are the same as in Fig.~\ref{sd_LR}.}
\label{sd_L}
\end{figure}

We compute the $\sigma_{p\chi}^{SD}$ and $\Phi_\mu$ 
in Fig.~\ref{sd_L} and Fig.~\ref{muflux_L} for $\tan\beta=~6$ and 30
in the LHLS and LMLS scenarios corresponding to the 
analyses of Fig.~\ref{MH_L} and Fig.~\ref{MM_L} respectively.  
The results show that the valid parameter space may be effectively 
probed in the future IceCube measurements. However, even the future IceCube reach
can not probe the Higgs resonance regions in these models since the corresponding
$\sigma_{p\chi}^{SD}$ and $\Phi_\mu$ values are too small.\\

\begin{figure}[!htbp]
\vspace*{-0.05in}
\mygraph{muflux_MHL}{muflux_MHL.eps}
\hspace*{0.5in}
\mygraph{muflux_MML}{muflux_MML.eps}
\caption{Plot of muon-flux $\Phi_\mu$ vs the mass of the LSP for 
two cases namely the LHLS (a) and LMLS (b) scenarios. Colours and conventions 
are the same as in Fig.~\ref{muflux_LR}.}
\label{muflux_L}
\end{figure}

\subsection{LHHS}
Considering the LHHS scenario associated with Fig.~\ref{LHHS} we compute 
$\sigma_{p\chi}^{SD}$ and $\Phi_\mu$ in Fig.~\ref{sd_hs} and 
Fig.~\ref{muflux_hs} with $\tan\beta$ = 30 for a higgsino 
dominated $\chonepm$. Only relic density and collider 
satisfied points along with the
$\gmin2$ constraint applied up to the 
level of 3$\sigma$ are shown. The present and future IceCube limits 
are shown in black and red lines respectively. Future IceCube experiment
can fully probe the allowed parameter points. However, the h-resonance region remains
beyond the reach of even the future IceCube bounds.\\

\begin{figure}[!htbp]
\vspace*{-0.05in}
\mygraph{sd_hs}{sd_hs.eps}
\hspace*{0.5in}
\mygraph{muflux_hs}{muflux_hs.eps}
\caption{Plot of SD direct detection cross-section and 
muon-flux $\Phi_\mu$ vs the mass of the LSP for the LHHS scenario with 
$\tan\beta$ = 30 for a higgsino dominated $\chonepm$.  
Only the points satisfying the relic density limits, collider 
constraints and the
$\gmin2$ constraint applied up to the 
level of 3$\sigma$ are shown. The present and future IceCube limits are shown in 
black and red lines respectively.}
\label{sd_and_muflux_hs}
\end{figure}

\section{New constraints on $\mgl$ and future search prospects}
\label{Sec:ConstraintsGluino}
In this section we consider a scenario where the gluinos, in addition to the 
EW sparticles, are relatively light. We choose several BPs from the LHLRS 
(Figs.~\ref{MH6LR} and \ref{MH30LR}), LHLS (Fig.~\ref{MH30L}) 
and LHHS (Fig.~\ref{LHHS}) models. Each BP captures 
the essential  features of the corresponding  model. 
We then  compute the new $\mgl$ limits  
in each case using the  ATLAS data for dedicated squark-gluino searches. 
The published data \cite{atlaspap1,atlaspap2,atlaspap3,atlaspap4,atlaspap5} 
are based on a series 
of conference reports \cite{atlas0l,atlas1l,atlas2l,atlas3b}. 
There are a few differences between the published results and the earlier
analyses so far as the details are concerned. However, there is no 
qualitative difference among the two sets of exclusion contours. Our 
analysis is based on Refs.~\cite{atlas0l,atlas1l,atlas2l,atlas3b}. 
This exercise also enables us to anticipate the probable signatures of 
different scenarios at future colliders. 
The characteristics of each BP are
briefly described in the following lines. The underlying sparticle spectra and the relevant BRs
are presented in Tables.~\ref{tab4}-\ref{tab6}. We like to mention that
for BP2, BP4 and BP6 we consider tilted scenarios with the slepton masses
driven closer to $\mchonepm$.  As already discussed in Secs. \ref{sec:LWLRS}
and \ref{sec:LHLRS}, constraints from slepton searches are weakened for tilted
scenarios. This allows the above-mentioned BPs which represent the Higgs 
resonance regions in the corresponding figures to become acceptable parameter points.

The point BP1 (BP2) is from the upper (lower) branch of the two 
regions satisfying the WMAP/PLANCK data in 
the LHLRS model (see Fig.~\ref{MH6LR}). From the BRs for BP1 in 
Table~\ref{tab5} and \ref{tab6}, 
it follows that in spite of the suppressed couplings of higgsino-dominated 
$\chonepm$, $\lsptwo$ and $\lspthree$ with appropriate fermion
-sfermion pairs of the first two generations, only the trilepton 
signal is potentially  viable for the LHC Run-II. Moreover, the  BR 
of the invisible mode $\lsptwo \ra \nu \wt{\nu}$ is quite large in this case. This is a generic 
feature of the branch under consideration.  Thus, $e^+ e^- \ra \met$ 
events are enhanced by $\lspone\lspone$, $\lsptwo\lsptwo$ and 
$\nu \wt{\nu}$ final states \cite{ad_virtual_lsp}. 
Such events can be searched for at the future $e^+ e^-$ colliders using the single photon tag from 
initial state radiation \cite{chen_drees}. In contrast, the invisible decays  
of $\lsptwo$ are rather suppressed in all other {\it higgsino models} studied in this paper.

\begin{table}[!htb]
{\small
\begin{center}
\begin{tabular}{|c|c|c|c|c|c|c|c|}
\hline
     & BP1 & BP2 & BP3 & BP4 & BP5 & BP6 & BP7 \\
\hline
Taken from & 1a & 1a & 1b & 1b &  3b & 3b & -- \\
           &    & (Tilted)  &  & (Tilted) &  & (Tilted) &  \\
\hline
$M_1$ & 256 & 59.5 & 202 & 62  & 227 & 62  &
60 \\
\hline
$M_2$ & 573 & 1000 & 522 & 1000 & 541 & 960 &
541 \\
\hline
$\mu$ & 286 & 500 & 261 & 500 & 270 & 480 &
271 \\
\hline
$\mlspone$ & 229 & 57 & 187 & 61 & 208 & 60 & 58 \\
\hline
$\mchonepm$ & 281 & 501 & 258 & 503 & 269 & 483 &
268 \\
\hline
$\mlsptwo$ & 295 & 503 & 269 & 503 &282 & 483 &
 270 \\
\hline
$\mlspthree$ & 303  & 510 & 272 & 511 & 282 & 490   & 282 \\
\hline
$M^D_{\tilde (e,\mu)_{L}}$ & 266 & 394 & 232 & 395 & 249 &
380 & 473 \\
\hline
$M^D_{\tilde (e,\mu)_{R}}$ & 265 & 394 & 232 & 395 & 2000 &
2000 & 473 \\
\hline
$\mstauone$ & 260 & 387 & 199 & 360 &  249 & 380 &
457 \\
\hline
$M^D_{\tilde \nu}$ & 255 & 386 & 219 & 388 & 236 &
372 & 466 \\
\hline
$\Omega_{\tilde \chi} h^2$ & 0.09 & 0.11 & 0.1 & 0.13 &  0.09 & 0.12 &  0.09 \\
\hline
$\sigma_{SI} (pb)$ $\times 10^{-9}$ & 15 & 0.13 & 5.4 & 0.03 & 7.5 & 0.03 &
0.2 \\
\hline
$a_\mu^{\rm SUSY}$ $\times 10^{-10}$ & 5.2 & 1.8 & 34 & 9.6 & 29 & 9.2 &
17 \\
\hline
\end{tabular}
  \end{center}
    \caption{The sparticle spectra corresponding to different BPs 
    chosen from  Fig.~\ref{MH_LR} and \ref{MH_L}. All masses are given in GeV. 
Only the closest figures are referred with the tilted scenarios, where the 
figures themselves represent regular scenarios with slepton masses at the 
midway between the masses of the LSP and $\chonepm$.}
\label{tab4}
}
\end{table}
BP2 is consistent with the slepton search data with a tilted slepton mass 
(see the discussions on the LWLRS$_{\chonepm}$ and LHLRS$_{\chonepm}$ models
in Secs. \ref{sec:LWLRS} and \ref{sec:LHLRS}). Here both $\lsptwo$ and $\lspthree$  
decay into $Z \lspone$ and $h \lspone$ while $\chonepm$ decay into $W^{\pm} \lspone$ 
modes with large BRs. Thus, the  $W h \lspone$ signal will strongly compete with the 
reduced trilepton channel during the next phase of the LHC experiments.  It may 
be noted that the current LHC constraints on the former signal in the {\it wino models}
\cite{atlasew1} are rather weak, even if the underlying decays are assumed to 
occur with 100\% BR. It is also worth recalling that this signal, in fact, 
does not look interesting in all the {\it wino models} except the LWHS model. 

\begin{table}[!htb]
{\small
\begin{center}
\hspace{-1 cm}
\begin{tabular}{|c|c|c|c|c|c|c|c|}
\hline
\hline
Decay   Modes                           &BP1    &BP2  &BP3 &BP4    &BP5     &BP6   &BP7     \\
\hline
$\gl    \ra \lspone  t \bar t  $      &6.6      &7    &3.5 &6     &4   &5.4    &2.6\\
$\quad    \ra \lsptwo  t \bar t  $      &17       &18   &12  &14    &14   &14      &14\\
$\quad    \ra \lspthree  t \bar t  $    &13       &16   &14  &12    &12   &13     &14\\
$\quad    \ra \lspfour  t \bar t  $     &1.6      &-    &1.7 &-     &1.5  &-    &1.5\\
$\quad    \ra \lspone  b \bar b  $      &1.2      &3    &2.3 &3     &3   &2.4    &1.4\\
$\quad    \ra \lsptwo  b \bar b  $      &-        &0.4  &4.4 &6.5   &5   &6.5      &5.4\\
$\quad    \ra \lspthree  b \bar b  $    &0.5      &0.3  &5   &6.3   &4   &6.2    &5\\
$\quad    \ra \lspfour  b \bar b  $     &2.7      &0.2  &2.6 &-   &2.5    &-      &2.4\\
$\quad  \ra  \chonepm t \bar b  $       &44       &52   	&42  	&50   &44   &50     &44\\
$\quad  \ra  \chtwopm t \bar b  $       &9.6      &-  	&9.4   	&-    &9.0	&-     &8\\
\hline
\hline
$\chonepm    \ra\snutau \tau       $       &32     &3     &37     &18     &69     &21.4   &- \\
$\quad    \ra\stau_{1} \nutau  $        &8      &1.5   &46     &15     &1.4    &0.4   &- \\
$\quad    \ra \stau_{2} \nutau $        &-      &-     &-      &2.6    &-      & -  &- \\
$\quad    \ra\snu_l l   $               &53     &4     &15     &2.6    &28     &3.6 &- \\
$\quad    \ra \wt{l}_L \nu_{l}$         &6.8    &1.2   &1.8    &0.6    &2.6    &0.8   &- \\
$\quad    \ra W^{\pm} \lspone$          &-      &90    &-      &61     &-      &74  &100\\
\hline
\hline

\end{tabular}
 \end{center}
  \caption{  Dominant decay modes and BRs of gluino and chargino for different BPs taken from 
  Fig.~\ref{MH_LR} and Fig.~\ref{MH_L}.  }
\label{tab5}
}
    \end{table}

The point BP3 (BP4), consistent with all the major constraints, is taken from the upper (lower) branch of the region
allowed by the WMAP/PLANCK data of Fig.~\ref{MH30LR}. The scenario represented by BP3 has a light $\stauone$. As a result,
$\chonepm$ - $\lsptwo$ pairs dominantly decay into final states involving multiple $\tau$s. Thus, searches 
with improved $\tau$-tagging may reveal this signature \cite{Bhattacharyya:2009cc}. However, 
a small but non-negligible trilepton signal can not be ruled out {\it a priori}. 
Thus, final states involving all of the three generations of leptons in 
different proportions, especially at an $e^+e^-$ collider, may be a hallmark of this model. 

BP4 similarly represents the generic features of the Higgs resonance region of Fig.~\ref{MH30LR}.
As in the previous case, final states with varieties of leptons belonging to 
all of the three generations is a feature of this scenario. However, the signal 
$W h \lspone$, if kinematically allowed, also looks promising for the LHC Run-II. 
Since the probability of $\tau$-rich final states is still sizable, observing 
both $\tau h \lspone$ and  $W h \lspone$ could be a smoking-gun signal. 
The former signature, though challenging at a hadron collider, is expected 
to be rather straightforward at an $e^+ e^-$ machine.

\begin{table}[!htb]
{\small
\begin{center}
\hspace{-1 cm}
\begin{tabular}{|c|c|c|c|c|c|c|c|}
\hline
\hline
Decay   Modes                           &BP1    &BP2  &BP3 &BP4    &BP5     &BP6   &BP7     \\
\hline
$\lsptwo   \ra  \slepl^\pm l^\mp       $&2      &4    &17.6     &2.64    &0.6   &3.6   &-    \\
$\quad  \ra\snu \bar{\nu}      $     & 57.6  &1.2     &0.5   &0.8     &18   &1.2  &-    \\
$\quad  \ra  \slepr^\pm l^\mp       $&10     &1.6    &22     &0.8     &    &-   & -   \\
$\quad  \ra  \stauonepm \tau^\mp    $&7.4    &3    &58       &28     &82   &20 & -   \\
$\quad \ra {{\wt\tau_2}^\pm}\tau^\mp$&22     &1.4    &1      &6      &-   &  & - \\
$\quad  \ra  \lspone h       $       &-      &65    &-       & 38    &-    &48   &62  \\
$\quad  \ra  \lspone Z      $        &-      &22     &-      & 22    & -    &28   &38 \\
\hline
\hline

$\lspthree  \ra  \slepl^\pm l^\mp       $&24       &-       &      &-    &48  & & -   \\
$\quad  \ra\snu \bar{\nu}      $     &4.8      &0.6     &8.4   &0.96 &6    &1.2 &  -  \\
$\quad  \ra  \slepr^\pm l^\mp       $&40       &0.4     &1.6   &0.4  &    &- &  -  \\
$\quad  \ra  \stauonepm \tau^\mp    $&19.2     &-       &86    &22   &46   &20 &  -  \\
$\quad \ra {{\wt\tau_2}^\pm}\tau^\mp$&12.8     &1.4     &4     &10   &-    &- &  -  \\
$\quad  \ra  \lspone h       $       &         &22      &-     &22   &-    &26    &20 \\
$\quad  \ra  \lspone Z      $        &-        &74      &-     &43   &-    &53   &80 \\
\hline

\end{tabular}
 \end{center}
  \caption{ Dominant decay modes and BRs of $\lsptwo$ and $\lspthree$ for different BPs taken from 
  Fig.~\ref{MH_LR} and Fig.~\ref{MH_L}.}
\label{tab6}

}
\end{table}

BP5 (BP6) captures the generic features of the upper (lower) branch of the 
region allowed by the WMAP/PLANCK data 
of Fig.~\ref{MH30L} quite well. Both the points are consistent with all the 
major constraints provided the slepton mass is tilted towards $\mchonepm$. 
BP5 also represents a scenario where all three generations of leptons may appear in the final 
state in different combinations, although a clear $\tau$-dominance is noticeable since the 
electroweakinos are higgsino-dominated. However, for BP6 the $\tau$-dominance is 
relatively mild. The $ W h \lspone$ signal, as and when sufficient luminosity 
accumulates, appears to be an attractive option.

BP7 is from Fig.~\ref{LHHS} and is consistent with all the major constraints. 
Here both the trilepton and the $W h \lspone$ events are expected to show up. 
However, it follows from Table~\ref{tab5} and \ref{tab6} that the relative rate of the later 
class of events are expected to be larger. As in the LWHS model, the LHHS model 
promises signatures other than the conventional trileptons 
from chargino-neutralino production. 

The gluino decay BRs for different scenarios are also presented in Table~\ref{tab5}. 
Since $\chonepm$, $\lsptwo$ and $\lspthree$ are higgsino-dominated, the final 
states with third generation of quarks overwhelm those containing leptons or light 
quark jets. Thus, it is expected that the search channels with tagged b-jets would 
yield the most stringent mass bounds on $\mgl$ from current data. It also 
follows that these are the best channels for gluino search in the {\it higgsino model} 
during the LHC Run-II. 

Before we compute the revised gluino mass limits in the {\it higgsino models},
we summarise the ATLAS SUSY searches sensitive to gluino pair production. 
After analysing the LHC Run-I ($\lum \sim$ 20 $\ifb$) data, the ATLAS  collaboration 
interpreted the results in the $n$-leptons + $m$-jets (with or without $b$ tagging) 
+ $\met$ channel with different integral values of $n$ and $m$ for various simplified models 
\cite{atlas0l,atlas1l,atlas2l,atlas3b}.  
For the inclusive jets + $0l$ + $\met$ channel,  depending on jet multiplicities, 
they defined five inclusive analyses channels (labelled as A to E) \cite{atlas0l}. 
The selection criteria used for 11 signal regions (SRs) are summarised  
in Table~1 of Ref. \cite{atlas0l}. 
In the absence of any significant excess, an upper limit on the number of events ($N_{BSM}$) 
from any Beyond the Standard Model (BSM) scenario was presented for various signal 
regions. The most effective signal regions for our analysis and the corresponding upper limits 
on $N_{BSM}$ for $\lum$ = 20.3 $\ifb$ are presented in 
Table~\ref{tab7}.

\begin{table}[!htb]
\begin{center}\
\begin{tabular}{||c||c|c||}
\hline
\hline
Channel	 	&Most effective  		&   Observed upper limits on	\\
	 	& signal Regions 		&   $N_{BSM}$ (95 $\%$ CL)	\\
\hline
\multirow{4}{*}{Jets $+ 0l + \met$\cite{atlas0l}}& SRC-Medium 	& 81.2		\\
				 								& SRD		& 15.5	\\
												& SRE-Medium	& 28.6	\\
												& SRE-Tight	& 8.3	\\
\hline
\multirow{2}{*}{Jets $+ 1l + \met$ \cite{atlas1l}}& Inclusive 6-jet($e$) & 4.6	\\
				 								& Inclusive 6-jet($\mu$)& 3.0	\\
\hline
Jets$ + 2SSl - 3l + \met$ \cite{atlas2l}			& SR3b	 & 3.9	\\
\hline
Jets (3b) $+ 0-1l + \met$ \cite{atlas3b} 			& SR-1l-6j-B& 3.0\\
\hline
\hline
       \end{tabular}\
       \end{center}
           \caption{The most effective signal regions for our analysis and the corresponding upper limits
on $N_{BSM}$ at 95 $\%$ CL  with $\lum$ = 20.3 $\ifb$  in the jets + $0l$ + $\met$ channel \cite {atlas0l}, 
jets + $1l$ + $\met$ channel \cite{atlas1l}, jets$ + 2SSl - 3l + \met$ channel\cite {atlas2l} 
and $0l$ +jets (3b) + $\met$ \cite{atlas3b} channel.}
\label{tab7}
\end{table}

We adopt the analysis of ``hard single-lepton" ($n = 1$) from Ref. \cite{atlas1l}.
In this channel, the ATLAS collaboration defined six inclusive and six binned 
signal regions treating electrons and  muons independently. Details of the signal regions 
are summarised in Table~4 of Ref.~\cite{atlas1l}. For our case, the most effective signal regions 
are  inclusive 6-jet (electron) and 6-jet (muon) (for the upper limits on $N_{BSM}$ in these two channels, 
see Table~\ref{tab7}). 

For the same sign (SS) dilepton analysis, the ATLAS collaboration 
considered either two isolated leptons ($e$ or $\mu$) with 
the same electric charge, or at least three isolated leptons ($3l$)\cite{atlas2l}. 
The \textit{SR3b} signal region yields the best limits for the {\it higgsino models}. In this
signal region, SS or $3l$ events are chosen with at least five jets and at least three b-jets. 
Corresponding upper limit on $N_{BSM}$ for $\lum$ = 20.3 $\ifb$  is 3.9 \cite{atlas2l}.  
Details of the selection criteria for the other signal regions are presented in Table 1 of \cite{atlas2l}. 

Next, we will briefly discuss the most important channel -
jets (at least 3 b-jets) + 0-1$l$ ($l$ = $e,\mu$) + $\met$ \cite{atlas3b}
which gives the most stringent bounds on $\mgl$ in the {\it higgsino models}. 
Selection criteria for the 9 signal regions are listed in Tables 1 and 2 of 
Ref. \cite{atlas3b} and upper limits on $N_{BSM}$ at 95 $\%$ confidence level 
(CL) are presented in Table 5 of Ref.~\cite{atlas3b}. The most effective signal region is 
$SR-1l-6j-C$\footnote{signal regions are classified as $A/B/C$ depending on 
$\met$ and $m_{eff}$.}, characterized by large $\met$ and at least 
six jets which includes at least three b-tagged jets. 
 
For electron, muon and jet identification, lepton-lepton isolation, 
lepton-jet isolation etc., we follow the ATLAS prescription 
as described in Refs.~\cite{atlas0l,atlas1l,atlas2l,atlas3b}. 
For b-tagging, we use the $P_T$ dependent b-tagging efficiencies 
presented by ATLAS collaboration in Ref. \cite{btagging}. 
After reconstruction of objects, we adopt all the signal 
regions defined by different selection criteria, introduced 
in Refs.~\cite{atlas0l,atlas1l,atlas2l,atlas3b}. 
For validation purpose, we also match the number of events 
and efficiencies of different cuts used for different signal regions 
in Refs.~\cite{atlas0l,atlas1l,atlas2l,atlas3b} with the ATLAS 
results. 

Using PYTHIA (v6.428)\cite{pythia} we generate the signal events 
in various channels from gluino pair production for the chosen 
BPs.  For the NLO $\gl \gl$ pair production cross-section calculation we 
use PROSPINO 2.1 \cite {prospino} with CTEQ6.6M PDF \cite {cteq6.6}. 
By comparing the simulated number of events with the corresponding 
upper limits on $N_{BSM}$ in the appropriate signal region, 
we calculate the new limits on $\mgl$ for different scenarios represented by 
BP1~-~BP7.

\begin{table}[!htb]
\begin{center}\
\begin{tabular}{||c||c||c||c||c||}
\hline
Points		& \multicolumn{4}{c|}{Limit on $\mgl$ (GeV)} 		\\
\cline{2-5}
& jets $+ 0l + \met$\cite{atlas0l} 	&jets $+ 1l + \met$ \cite{atlas1l}	& jets $+ 2l + \met$ \cite{atlas2l}	&$0l$ +jets (3b) + $\met$ \cite{atlas3b}	\\
\hline
BP1	 	& 675 	 		&1125 	 		& 1250			& 1340		\\
\hline
BP2	 	& 	1080 		&	1175 		&1135 			&1360		\\
\hline
BP3	 	& 815	 		&	 1100		&1180 			&1320		\\
\hline
BP4	 	& 1050	 		&	1160 		&1135 			&1345		\\
\hline
BP5	 	& 770	 		&1105	 		&1210 			&1330		\\
\hline
BP6	 	& 1075	 		&1160	 		& 1135			&1345		\\
\hline
BP7	 	& 980	 		&1130	 		&1135 			&1325		\\
\hline
\hline
       \end{tabular}\
       \end{center}
           \caption{Limits on $\mgl$ for different BPs
           using the ATLAS jets + $0l$ + $\met$ data\cite {atlas0l}, 
           jets + $1l$ + $\met$ data\cite{atlas1l}, 
           jets $+ 2l + \met$ (SSD) data\cite {atlas2l} 
           and $0l$ +jets(3b) + $\met$ \cite{atlas3b} data.}
\label{tab8}
          \end{table}


The revised limits on $\mgl$ in different {\it higgsino models} are presented in Table~\ref{tab8}. 
As expected, the strongest limits come from search channels involving tagged b-jets.
The limits are practically independent of the choice of the slepton masses. 
It may be recalled that in the {\it wino model} the best limits come from  the jets$ + 1 l + \met$ channel. 
This table also illustrates the importance of multichannel search for the {\it higgsino model}. 
The size of jets$ + 0 l + \met$ signal can potentially distinguish some of the
{\it higgsino models} from the others.  However, our results are based on 
the generic strategies for squark-gluino searches devised by the ATLAS 
collaboration using tagged b-jets. In the LHC Run-II, more dedicated searches, e.g.,
the detection of a Higgs boson in a gluino decay cascade may provide more definite 
information on the underlying {\it higgsino model}. 

\section{Conclusion}
\label{Sec:Conclusion}
The focus of this paper is on the phenomenology of the
{\it higgsino} and the {\it mixed models} of the electroweakinos
 and to compare and contrast them with that of the corresponding 
{\it wino models} studied in Ref.~\cite{older}. 
In this concluding section we summarize our main results in 
the light of the three major constraints (see Table~\ref{summarytab2}).

\begin{table}[!htb]
\small{
\begin{center}
\begin{tabular}{|c||c|c|c|}
\hline
Model    & Trilepton Constraint  & Processes Leading to Correct          & The $\gmin2$ Constraint  \\
         &                  & Relic density                         &     Satisfied at        \\
\hline
\multirow{3}{*}{LHLRS}     &            & $\lspone \lspone \ra W^{+} W^{-}, 
t \bar t, ZZ, Zh$  & (a) 3$\sigma$ (low $\tan\beta$) \\
                           &  Degraded w.r.t. the LWLRS case    & Z/h resonance annihilation
& (b) $\le 2\sigma$ (high $\tan\beta$) \\
                           &  & $\stauone$ coannihilation 
(only for high $\tan\beta$)  &  \\
\hline
\multirow{2}{*}{LHLS}      & Degraded w.r.t. the LWLS case & $\lspone \lspone \ra W^{+} W^{-}, t \bar t, ZZ$  
& Same as above \\
                           &                           & Z/h resonance annihilation   & \\
\hline
\multirow{2}{*}{LHHS} & Degraded w.r.t. the LWHS case. & $\lspone \lspone
 \ra W^{+} W^{-}, ZZ, t \bar t$  & $\le 2\sigma$ for high $\tan\beta$\\
                           &    & Z/h resonance annihilation      &  \\
\hline
\multirow{3}{*}{LMLRS}     & Degraded w.r.t. the LWLRS case         & $\lspone \lspone \ra W^{+} W^{-}, 
t \bar t $  & (a) 3$\sigma$ (low $\tan\beta$)\\
                           & but stronger than the LHLRS model.    & Z/h resonance annihilation
& (b) $\le 2\sigma$ (high $\tan\beta$)\\
                           &   & $\stauone$ coannihilation 
(only for high $\tan\beta$)  &   \\
\hline
\multirow{2}{*}{LMLS}     & Degraded w.r.t. the LWLS case.  & $\lspone \lspone \ra W^{+} W^{-}, t \bar t, ZZ$  
& Same as above\\
              & but stronger than the LHLS model & Z/h resonance annihilation  & \\
\hline
\end{tabular}
\end{center}
\caption{Summary of the impact of   
the three major constraints on the models analyzed in this work. 
Here, LWLS corresponds to the Light 
Wino and Light Left Sleptons model. Similarly, LWLRS refers to the Light Wino and 
Light Left and Right Sleptons and 
LWHS corresponds to the Light Wino and 
Heavy Sleptons scenarios respectively.  These {\it wino models} were discussed 
in detail in Ref.\cite{older}.}
\label{summarytab2}
}
\end{table}

To give the readers some feelings for the numerical values 
of the revised LHC mass limits we note that in the LHLRS (for low 
$\tan\beta$) and LHLS (for both values of $\tan\beta$) models the 
lower bounds on $\mchonepm$ are 380 GeV and 360 GeV respectively. 
These bounds are significantly weaker than the similar bounds in the 
corresponding {\it wino models} which have the ballpark values of 
$\sim 600$ GeV. It is interesting to note that the entire exclusion contour 
in the LHLRS model with high $\tan\beta$ is superseded by the theoretical 
constraints. However, for small LSP masses, the bounds from the slepton 
searches translate into stronger bounds: $\mchonepm$ $\geq$ 650 GeV (LHLRS) 
and 600 GeV (LHLS). As discussed in detail in the text these bounds get 
weaker in the tilted LHLRS$_{\chonepm}$  and LHLS$_{\chonepm}$ models 
yielding $\mchonepm$ $\geq$ 450 GeV (LHLRS) and 410 (LHLS) GeV. 
For higher LSP masses both the bounds obtained directly from 
the trilepton searches and those deduced from the constraints in 
the slepton sector become relaxed and eventually disappear for 
certain LSP masses which for each model can easily be read off 
from the figures concerned. In the LHHS model the bound is $\mchonepm$ 
$\geq$ 175 GeV for negligible LSP masses. This is rather weak even 
in comparison with the corresponding limit in the LWHS model 
which is the most relaxed limit among the {\it wino models}.
We have also considered the LHC constraints in the {\it mixed models}. 
As expected, the LHC limits lie in between the corresponding 
ones in the {\it wino} 
and the {\it higgsino models}. However, no qualitatively 
new feature emerges from this analysis.

In all models considered in this paper and in Ref \cite{older} 
with low $\tan\beta$ the predictions for $\gmin2$ are consistent 
with the data only at the level of 3$\sigma$ after imposing the 
LHC constraints. The constraint can be more effective 
only in high $\tan\beta$ scenarios. 

In the parameter spaces of the {\it higgsino} and the 
{\it mixed models} there are two distinct branches allowed by the 
WMAP/PLANCK constraints, as in the {\it wino models}. In the upper 
branches the LSP pair annihilations into various channels turn out to 
be the dominant DM relic density producing mechanism as illustrated 
in Table \ref{summarytab2}. Only in the LHLRS and the 
LMLRS models with large $\tan\beta$, 
the LSP-stau coannihilation is important. In contrast, the upper branches 
in the {\it wino models} are dominated by different coannihilation processes.

In the lower branches of the {\it higgsino} and the {\it mixed models} 
analyzed in this work, the DM production is mainly due to 
LSP pair annihilation 
via the h-resonance.  The APSs are larger in models with tilted slepton masses 
for reasons explained in the text. In contrast, this mechanism is generically 
under pressure in the {\it wino models} with heavy sleptons either due to 
the above tension with the $\gmin2$ constraint or due to the LHC constraints or both.

The inclusion of the WMAP/PLANCK and the $\gmin2$ constraints 
in our analysis severely restricts the APSs by imposing both upper and 
lower mass bounds in most of the models studied here. 
At high $\tan\beta$, most of the {\it wino}, {\it mixed} and the {\it higgsino models} 
have narrow APSs surviving all the major constraints.
We now summarize our main findings regarding the prospects of having novel 
signatures at the LHC Run-II and at the ILC. 

In the {\it higgsino models}
$\lsptwo$ or $\lspthree$ decaying into $Z h \lspone$ with 
large BR are rather common (see the examples in  Table~\ref{tab6}). 
They occur even if the sleptons are lighter than the electroweakinos. In 
contrast, these decays occur with large BRs in the {\it wino models} with  
heavy sleptons provided the DM constraints are relaxed. The discovery of
light sleptons together with the observation of the
$W h \met$ events due to chargino-neutralino production during the LHC Run-II 
could be the hallmark of the {\it higgsino models}. 
If we focus on this signal in parameter spaces consistent 
with the $\gmin2$ and the DM relic density constraints, 
then only zones with high $\tan\beta$ and DM relic density 
production via LSP pair annihilation into the h-resonance are 
acceptable. On the other hand this  DM producing mechanism is 
generically disfavoured in the {\it wino models} with high $\tan\beta$
 as noted earlier.

In some
regions of the APSs, especially in the upper branches of the 
regions allowed by the WMAP/PLANCK data in the $\mchonepm-\mlspone$ plane, 
the conventional 
trilepton channel appears to be the best bet in many models 
considered in this paper and in Ref. \cite{older}. With improved 
$\tau$-tagging efficiencies, the final states with multiple $\tau$'s 
as analysed in Ref. \cite{Bhattacharyya:2009cc} for models with high $\tan\beta$ 
may provide alternative/complementary
search channels during the LHC Run-II. Several novel signatures which could be relevant at the 
ILC have also been discussed in Sec. \ref{Sec:ConstraintsGluino}. 
In particular, the comparison of the constraints obtained in this paper and the ones in Ref.
\cite{older} clearly indicates that if the ILC indeed operates at around 500~GeV during its first run as planned, 
then the electroweakinos in the {\it higgsino model} have  larger probabilities 
of being within its striking range (see, for e.g., Fig.~\ref{LHHS}).

Assuming the gluinos to be light in addition to the electroweak sparticles while all 
squarks are heavy, we have revisited the gluino mass limits in Sec. \ref{Sec:ConstraintsGluino}
using the ATLAS data.  
As in the {\it wino model}, this analysis emphasizes the importance of multichannel searches. 
It follows that the conventional jets + 0 $l$ + $\met$ signal has the poorest sensitivity 
in a wide variety of the {\it higgsino models} (see the results in 
Table~\ref{tab8} for different BPs  in Sec. \ref{Sec:ConstraintsGluino}), whereas 
jets + 1$l$ or 2$l$ + $\met$ signal has a better sensitivity. The best channel for
probing the {\it higgsino models} involves multiple tagged b-jets.
The gluino mass limits are stronger in general than the ones obtained in the corresponding 
{\it wino models}.
This observation may help to formulate the future strategies for gluino searches 
in the context of the {\it higgsino models} and distinguish between
the {\it wino} and the {\it higgsino models}, if a signal is seen. 
Moreover, one can distinguish among various {\it higgsino models} 
consistent with the major constraints by the relative rates of jets + 0 $l$ + $\met$ 
and jets (3b)+ 0 $l$ + $\met$ events (see Table \ref{tab7}).

We next summarize the prospects of direct and indirect detection of DM in the context of
the {\it higgsino} models. 
There is a significant bino-higgsino mixing in the LSP in the {\it higgsino models}
especially in the portions of the parameter spaces consistent with the DM relic 
density constraint and characterized by relatively small 
$\mchonepm -\mlspone$. The spin-independent
direct detection cross-section $\sigma_{\tilde \chi p}^{\rm SI}$ is larger in all such
cases compared to the corresponding {\it wino models}. Although the cross-sections exceed the LUX limits in most of the cases, they stay within an order of magnitude of the same limits.    
However, 
the cross-section for the points from the Higgs resonance region 
allowed by the major constraints
(e.g. in Figs.~\ref{dd_MMLR}, \ref{dd_L} and \ref{dd_HS}) lie well below the LUX limit. 
All the models can be probed by the future XENON1T experiment 
irrespective of the degree of uncertainties in 
$\sigma_{\tilde \chi p}^{\rm SI}$.   

The spin-dependent cross-section
$\sigma_{\tilde \chi p}^{\rm SD}$ and muon flux values for the neutrino 
signals 
in the {\it higgsino} models are enhanced
compared to their {\it wino} model counterparts.
Most of the points in almost all the scenarios are allowed by the present
IceCube data. In some cases the situation seems very interesting
since the values of $\sigma_{\tilde \chi p}^{\rm SD}$ and muon flux lie
very close to the present experimental bound
(see e.g. Figs. \ref{sd_MHLR}, \ref{muflux_MHLR},
\ref{sd_MHL}, \ref{muflux_MHL} and \ref{sd_and_muflux_hs}).
All the other cases would decisively be 
probed by the future IceCube searches.
For the points representing the $h$-resonance
region in Figs. \ref{sd_MMLR}, \ref{muflux_MMLR}, \ref{sd_L},
\ref{muflux_L} and \ref{sd_and_muflux_hs} the values of these
observables are too small to be detected even by the future 
IceCube reach.

{ \bf Acknowledgments : } 
MC would like to thank Council of Scientific and Industrial Research, Government of 
India for financial support. The work of AC was partially supported by funding available 
from the Department of Atomic Energy, Government of India for the Regional Centre for 
Accelerator-based Particle Physics (RECAPP), Harish-Chandra Research Institute. 
AD acknowledges the award of a Senior Scientist position by the Indian National Science 
Academy.


\end{document}